\documentclass[aps,pra,reprint,twocolumn,showpacs,floatfix,superscriptaddress]{revtex4-1}
\usepackage{amssymb,amsmath,amstext}
\usepackage[utf8]{inputenc}
\usepackage[OT4]{fontenc}
\usepackage{graphicx}
\usepackage{bm}
\usepackage{hyperref}
\usepackage{xcolor}

\usepackage{physics}
\usepackage{ulem}

\begin{document}


\title{Optical lattice for tripod-like atomic level structure}

\author{Piotr Kubala}
\affiliation{Institute of Theoretical Physics, Jagiellonian University in Krak\'ow,  \L{}ojasiewicza 11, 30-348 Krak\'ow, Poland }
\author{Jakub Zakrzewski}
\affiliation{Institute of Theoretical Physics, Jagiellonian University in Krak\'ow,  \L{}ojasiewicza 11, 30-348 Krak\'ow, Poland }
\affiliation{Mark Kac Complex
Systems Research Center, Jagiellonian University in Krakow, Krak\'ow,
Poland. }

\author{Mateusz Łącki}
\affiliation{Institute of Theoretical Physics, Jagiellonian University in Krak\'ow,  \L{}ojasiewicza 11, 30-348 Krak\'ow, Poland }
\email{mateusz.lacki@uj.edu.pl}

\begin{abstract}
Standard optical potentials use  off-resonant laser standing wave induced AC-Stark shift. In a recent development
[Phys. Rev. Lett. {\bf 117}, 233001 (2016)] a three-level scheme in $\Lambda$ configuration coupled coherently by resonant laser fields was introduced leading to an effective lattice with subwavelength potential peaks. Here, as an extension of that work, a four level atomic setup in the tripod configuration is used to create spin $1/2$-like  two-dimensional dark-space with 1D motion and the presence of external gauge fields. Most interestingly for possible applications, the lifetime for a dark subspace motion is up to two orders of magnitude larger than for a similar $\Lambda$ system. The model is quite flexible leading to lattices with significant nearest, next-nearest, or next-next-nearest hopping amplitudes, $J_1,J_2,J_3$ opening up new intriguing possibilities to study, e.g.  frustrated systems. The characteristic Wannier functions lead also to a new type of inter-site interactions not realizable in typical optical lattices.

\end{abstract}

\date{\today}

\maketitle

\section{Introduction}

AC-Stark shift based optical potentials induced by  far-detuned laser standing waves  has enabled to implement  discrete lattice models~\cite{Jaksch1998} linking ultra-cold atomic physics with condensed matter physics. Or rather enriching the latter with bosonic systems  such as e.g. Bose-Hubbard (BH) model. The experimental demonstration of a quantum phase transition between the superfluid and Mott insulating phases~\cite{Greiner2002}  was followed by intensive investigations in different, more complex schemes~\cite{Lewenstein2012,Dutta15,Spielman2019} involving  spinor lattice gases, long range interactions, disordered systems, or an implementation of topological insulators. 

The standing wave optical potentials proved to be very versatile allowing to create, typically with the application of additional Raman lasers, interesting coupling between sites, e.g., leading to a construction of artificial gauge fields or spin-orbit coupling as reviewed in \cite{Spielman2019}. The atomic ground state sublevels could serve as an additional synthetic dimension \cite{Boada12,Celi14,Suszalski16}, allowing, e.g. to extend the Hall physics to four dimensions \cite{Lohse18}. Still  the standing wave optical potential has some drawbacks. The typical
$\cos^2(k_Lx)$ spatial dependence (with $k_L$ being the laser light wavevector) leads to the dominance of nearest neighbor tunneling over hops involving separated sites.  Similarly on-site interactions dominate over inter-site terms making investigations of interaction related physics for spinless fermions in optical lattices difficult.

Recently  an alternative scheme for creating optical potentials has been proposed~\cite{Lacki2016,Gorshkov2016}.  It relies on a {\it resonant} dipole-coupling of three atomic levels with the position-dependent Rabi frequencies involving a common atomic excited state. This differs from the standard approach where two-photon resonant lasers are far detuned from a single-photon transition \cite{Larson2008,Larson2009}. The resulting Lambda system is characterized by a position-dependent dark state. The dynamics of atoms constrained to the dark state is that of  a particle moving in the presence of a scalar potential which features evenly spaced subwavelength peaks. Early ${}^{171}\textrm{Yb}$ experiments~\cite{Wang2018,Rolston2020} confirmed the expected band structure,  but the system lifetimes were, {disappointingly}, at least one order of magnitude lower than for AC-Stark potentials.

In this work we present the tripod system \cite{Ruseckas2005,Ohberg2011}, with four resonantly coupled levels that is an interesting extension of the $\Lambda$ system. It features two degenerate dark states implementing spin $1/2$-like physics {providing at the same time a new possible realization of spin-orbit coupling in a one-dimensional lattice supplementing the existing schemes \cite{Lin11,Hamner15}}. In Section \ref{sec:Hamiltonian} we adapt the derivation of the $\Lambda$ system dark-state description to the tripod scheme.  {We detail the resulting periodic spin $1/2$-like model for the movement of the particle in the gauge field and discuss the Bloch theory including lifetime computation of the dark state bands}. The tight-binding description of atoms populating  {low lying} bands is discussed in Section \ref{subsec:bands:tightBinding}.
As it turns out, the model leads, in a natural way, to a quite peculiar extended Hubbard model with significant hopping not only to the nearest sites (nn) but also to the next nearest neighbors (nnn) as well as to next-next nearest neighbors (nnnn). Such a highly interesting and unusual property is due to the shape of the Wannier basis functions corresponding to the nonstandard lattice felt by the atoms. This opens up a possibility of frustration related studies in the model. The conclusions are drawn and future perspectives involving the study of interacting particles are discussed in Sec. \ref{sec:summary}.

\section{The Hamiltonian}
\label{sec:Hamiltonian}

We consider a gas of ultracold atoms whose motion is restricted to one dimension for example by a strong external optical potential of the form $V(y,z)=m_a\omega^2(y^2+z^2)/2$. The atoms populate three ground state configuration atomic states $|g_1\rangle,|g_2\rangle,|g_3\rangle$ that are coupled to an excited state $\ket{e}$.  The dipole coupling of each of $\ket{g_i}$  to $\ket{e}$ is  {characterized} by a Rabi frequency $\Omega_i(x)$. The wavelength of the three lasers is assumed to be  equal to $\lambda_L$. In a rotating frame, after neglecting rapidly oscillating terms, the Hamiltonian {of the system considered} reads
\begin{equation} \label{eq:ham_bare}
    H = -\frac{\hbar^2}{2m_a} \pdv[2]{x} + H_{a}(x),
\end{equation}
where $m_a$ is the {atomic} mass and 
\begin{equation}\label{eq:ham_at_bare}
H_{a}(x)=\begin{pmatrix}
-\Delta - i\Gamma_e/2 & \Omega_1^*(x) & \Omega_2^*(x) & \Omega_3^*(x)\\
\Omega_1(x) & 0 & 0 & 0 \\
\Omega_2(x) & 0 & 0 & 0 \\
\Omega_3(x) & 0 & 0 & 0
\end{pmatrix}.
\end{equation}
Here $\Delta$ is a possible common detuning of all three lasers and $\Gamma_e$ is  spontaneous emission rate of the excited state. 
We  consider Rabi frequencies of the form \begin{eqnarray}
\Omega_1(x)&=&\Omega_1\sin(k_Lx),\nonumber\\
\Omega_2(x)&=&\Omega_2 \sin(k_Lx+a), \nonumber\\
\Omega_3(x)&=&\Omega_3,\label{eq:Omegas}
\end{eqnarray} 
where $k_L=2\pi/\lambda_L$.  The $\lambda_L$-periodicity defines a natural energy scale -- the recoil energy equal to $E_R={\hbar^2k_L^2}/{(2m_a)}.$

The matrix {$H_a(x)$}, in Eq.~\eqref{eq:ham_at_bare}, for $\Gamma_e\neq 0$, is non-Hermitian. It is diagonalized by finding a biorthogonal set of right and left eigenvectors. When using bra/ket notation, the bra vector always refers to a proper left eigenvector, a part of biorthogonal set. There exist two linearly independent right eigenvectors $\ket{D_1(x)}$, $\ket{D_2(x)}$ corresponding to energy  $E(x)=0$  -- two dark states. They are of the form
\begin{equation}\label{eq:Omegauperp}
|D_i(x)\rangle= \sum_{n=1}^3 u_{i,n}\ket{g_i},\quad \vb{\Omega}(x)\perp \vb{u}_{i}, 
\end{equation}
They are not affected by value of $\Delta,\Gamma_e$.

\begin{figure}[htb]
    \centering
    \includegraphics[width=0.9\linewidth]{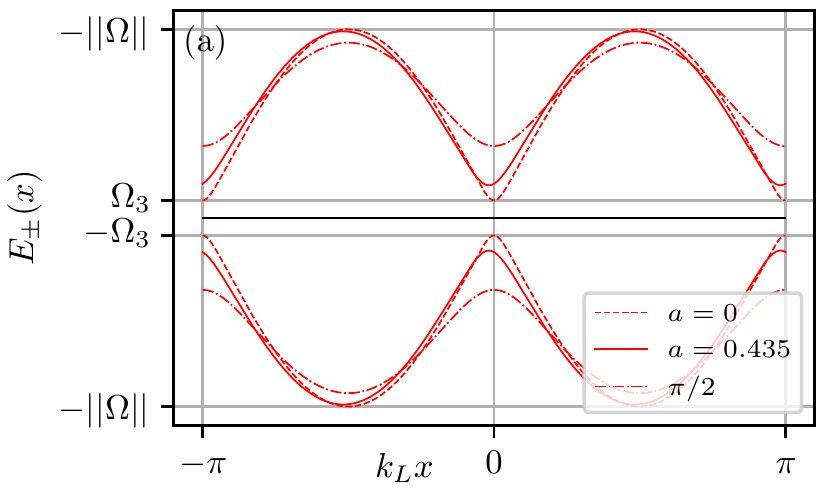}
    \caption{The diagonal energies of bright channels $E_+(x)>0$, $E_-(x)<0$ for $\Omega_1:\Omega_2:\Omega_3=5000:2000:500$. For $a\in \{0,\pi/2\}$ the $E_+(x)$ has minimum for $x=0$. 
    }
    \label{fig:tripodEpEm}
\end{figure}

The Hamiltonian matrix \eqref{eq:ham_at_bare} has also two (right) bright states $\ket{B_\pm(x)}$ with energies
\begin{equation} \label{eq:bright_e}
    E_\pm(x) = -\tilde{\Delta}/2 \pm 
    \sqrt{(\tilde{\Delta}/2)^2 + \norm{\vb{\Omega}(x)}^2},
\end{equation} 
where $\tilde{\Delta} = \Delta + i \Gamma / 2$.  These states read 
\begin{equation} \label{eq:bright_v}
    \ket{B_\pm(x)}= \mathcal{N}_\pm(x) \left(E_\pm(x)|e\rangle + \sum_{i=1}^3 {\Omega}_i(x)|g_i\rangle\right),
\end{equation}
where
\begin{align}
\mathcal{N}_\pm(x)&=\frac{1}{\sqrt{(E_\pm(x))^2+||\vb{\Omega}||^2}},\nonumber \\ 
||\vb{\Omega}||&=\sqrt{|\Omega_1(x)|^2+|\Omega_2(x)|^2+|\Omega_3(x)|^2}
\end{align}
are the normalization factors. The left eigenvectors, that together with \eqref{eq:bright_v} form a biorthonormal set, are given by:
 
\begin{equation} \label{eq:bright_v_left}
    \bra{B_\pm(x)}= \mathcal{N}_\pm(x) \left(E_\pm(x)\langle e| + \sum_{i=1}^3 {\Omega}_i^*(x)\langle g_i|\right).
\end{equation}

 For $\Gamma_e\neq 0$ the states $|B_\pm(x)\rangle$ undergo a spontaneous emission with the rate comparable to that of an excited state $|e\rangle$. In this work we are interested in the case when  atoms populate primarily the stable channels $|D_1(x)\rangle,D_2(x)\rangle$ and the energy scale set by  $\Omega_i$ dominates the kinetic energy of atoms.
 This also ensures that a phenomenological description of losses from largely unpopulated $|e\rangle$ via $i\Gamma_e$ is justified  $\Gamma_e\ll ||\mathbf{\Omega}||$ \footnote{In general, the more complete treatment of losses would be by the Lindblad master equation approach.}

The gap $\Delta E$ between the dark and the bright state channels, for $\Delta=0$, is given by $\min E_+(x).$ For $a=0$ we have $\Delta E=|\Omega_3|$. For $a>0$ the gap $\Delta E$ increases until $a=\pi/2$ where $\Delta E=\min\{\sqrt{\Omega_3^2+\Omega_2^2},\sqrt{\Omega_3^2+\Omega_1^2}\}$. The dependence of $E_\pm$ on x for few selected values of $a$ is shown in Fig.\ref{fig:tripodEpEm}.

It is worth stressing that we  focus on non-interacting bosons or physics of ultracold spinless fermions, where direct collisions are suppressed. If this is not the case one may wonder whether the collision of two atoms in dark state may not lead to one particle in each bright state as $E_+(x)+E_-(x)=-\tilde\Delta$. We leave this question open for the interacting case study, here 
let us mention that the process may be suppressed by taking a sufficient detuning $\Delta$ still in the limit $\Delta \lessapprox ||\mathbf{\Omega}||$.

The Hamiltonian~\eqref{eq:ham_bare} may be  addressed using a Born-Oppenheimer type transformation~\cite{Ruseckas2005,Lacki2016} applying
\begin{eqnarray}
{\cal{B}}(x)=\{\ket{D_1(x)},\ket{D_2(x)},\ket{B_+(x)},\ket{B_-(x)}\}
\end{eqnarray}
as a position-dependent basis. Writing an arbitrary wavefunction in this basis as: 
\begin{align}
|\psi(x)\rangle =d_1(x) \ket{D_1(x)}&+d_2(x) \ket{D_2(x)}+\nonumber \\
& +b_-(x) \ket{B_-(x)}+b_+(x) \ket{B_+(x)},
\label{eq:exp}
\end{align}
yields the following  Hamiltonian matrix:
\begin{equation} \label{eq:ham_dressed}
    H_{\cal{B}} = \frac{[P-A(x)]^2}{2m_a} + 
    \textrm{diag}[0,0, E_+(x), E_-(x)],
\end{equation}
or
\begin{align} \label{eq:ham_dressed2}
    H_{\cal{B}} = \frac{1}{2m_a}&\left[{P^2} -2A(x)P + \Phi(x) \right]+\nonumber \\
    &+\textrm{diag}[0,0, E_+(x), E_-(x)].
\end{align}
The $H_{\cal{B}}$ now acts on vectors of the form:
\begin{align}
    \psi(x)\equiv \left(\begin{array}{c} d_1(x) \\ d_2(x) \\ b_+(x) \\ b_-(x) \end{array}\right).
\end{align}
The operator $P=-i\hbar \partial_x \otimes \mathbf{1}_4$. The $A(x)$ is given by
\begin{equation} \label{eq:A}
    A_{MN}(x) = i\hbar \bra{M(x)} \partial_x \ket{N(x)},\quad M(x),N(x)\in {\cal{B}}(x).
\end{equation}
and
\begin{equation}
    \Phi(x)=A(x)^2 + i\hbar \partial_x A(x).
\end{equation}

\subsection{The dark state subspace}

The large diagonal terms $E_-(x),E_+(x)$ in Eq.~\eqref{eq:ham_dressed2} allow for the separation of the dark-state physics in Hamiltonian \eqref{eq:ham_dressed2} by the following dark-state projection:

\begin{equation}
H_2=QH_{\cal{B}}Q,\ \ Q=\ketbra{D_1(x)}{D_1(x)} + \ketbra{D_2(x)}{D_2(x)}
\label{eq:H2}
\end{equation}
This will be evident from the upcoming numerical analysis.

The states $\ket{D_1(x)},\ket{D_2(x)}$ are energy-degenerate and thus not uniquely defined. Different choices of basis lead to equivalent description of the model. We favor those leading to simple, well-behaved and intuitive potentials $A(x)$. We consider only $\lambda_L$-periodic $|D_i(x)\rangle$. We also opt to work with $|D_i(x)\rangle$ with real coefficients, which automatically implies $A_{11}(x)=A_{22}(x)=0$.  

Let us take 

\begin{equation} 
|{D}_1(x)\rangle \propto |\xi\rangle \cross \vb{\Omega}(x),
\label{eq:genD1}
\end{equation}
and
\begin{equation} 
|{D}_2(x)\rangle\propto |{D}_1(x)\rangle \cross \vb{\Omega}(x).
\label{eq:genD2}
\end{equation} 
This assures their mutual orthogonality and by Eq.~\eqref{eq:Omegauperp} such vectors are indeed dark. The vector $|\xi\rangle$ cannot be parallel to $\mathbf{\Omega}(x)$ but otherwise can be arbitrary. We choose:
\begin{equation}
    |\xi\rangle =\Omega_1 |g_1\rangle + \Omega_2 |g_2\rangle.
    \label{eq:vxi}
\end{equation}
We will later discuss the advantages of the above choice for $|\xi\rangle$, namely good analytic properties in the limit $a\to 0$.  Using Eq.~\eqref{eq:vxi} we find
\begin{equation}
    |D_1(x)\rangle=N_1(x)\left(
    \begin{array}{c}
     \Omega_2 \Omega_3 \\
     -\Omega_1 \Omega_3 \\ 
     \Omega_1 \Omega_2[ \sin(a+k_Lx)-\sin(k_Lx)]
    \end{array}
    \right)
\label{eq:bestD1}
\end{equation}
and 
\begin{equation}
|D_2(x)\rangle\!=\!N_2(x)\!\left(\!\!
\begin{array}{c}
 {\Omega_1} \left[\Omega_2^2 f(k_Lx)+\Omega_3^2\right] \\
 {\Omega_2} \left[\Omega_1^2 f(\pi/2+k_L x-a/2)+\Omega_3^2\right] \\
 -{\Omega_3} \left[\Omega_2^2 \sin (a\!+\!k_L x)+\Omega_1^2 \sin (k_Lx)\right] \\
\end{array}
\right),
\label{eq:bestD2}
\end{equation}
where
\begin{equation}
    f(y)=\sin (a+y)[\sin (a+y)-\sin(y)].
\end{equation}
When $a=0$ the state $|D_1(x)\rangle$ becomes position-independent 
\begin{equation}
|D_1(x)\rangle =\frac{\Omega_2}{\sqrt{\Omega_1^2+\Omega_2^2}}|g_1\rangle -\frac{\Omega_1}{\sqrt{\Omega_1^2+\Omega_2^2}}|g_2\rangle.
\label{eq:darkD1indep}
\end{equation} and
\begin{equation}
D_2(x)=N_2(x)[\Omega_1\Omega_3|g_1\rangle+\Omega_2\Omega_3|g_2\rangle-
(\Omega_1^2+\Omega_2^2)\sin(k_Lx)|g_3\rangle]. \label{eq:darkD2indep}
\end{equation}
The latter can be written as 
\begin{equation}
    |D_2(x)\rangle = \frac{1}{\sqrt{\Omega_p^2+\Omega_c^2}}[\Omega_p |a\rangle - \Omega_c\sin(k_L x) |g_3\rangle]
\end{equation}
where $|a\rangle = \cos\beta |g_1\rangle +\sin\beta |g_2\rangle, \tan\beta=\Omega_2/\Omega_1, \Omega_c=\sqrt{\Omega_1^2+\Omega_2^2}, \Omega_p=\Omega_3$. The above form is formally identical to the form of a single dark state in the $\Lambda$ system configuration~\cite{Lacki2016}.

\begin{figure}[ht!]
    \centering
    \includegraphics[width=0.9\linewidth]{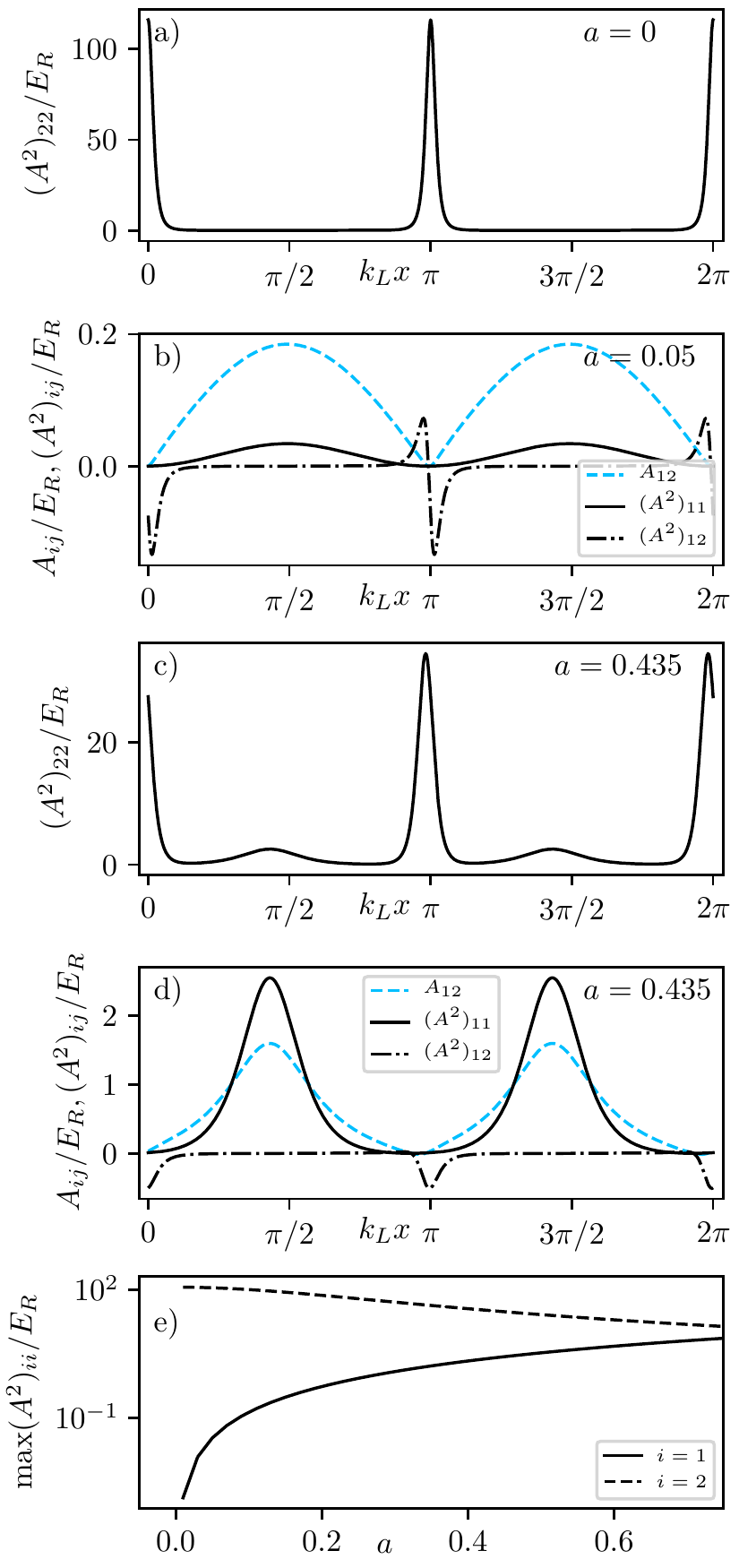}
    \caption{The spatial dependence of elements of matrices $A(x)$ and $A^2(x)$ describing the couplings within the dark state subspace for $\Omega_1:\Omega_2:\Omega_3=50:20:5$. The value of the phase shift $a$ is indicated in each of the panels. Panel a) shows the limiting case of $a=0$ for $(A^2)_{22}(x)$ [see also Eq.~\eqref{eq:A22Vna}]. Panel b) shows $A_{12}(x), (A^2)_{11}$ and $(A^2)_{12}(x)$ for a small, but nonzero $a=0.05$. Coefficient $(A^2)_{22}(x)$ for this $a$ is similar to one in a). Panel c) shows $(A^2)_{22}(x)$ for a larger $a=0.435$, while smaller coefficients: $A_{12}(x), (A^2)_{11}$ and $(A^2)_{12}(x)$ are depicted on a separate panel d).  The dependence of peak height of $(A^2)_{11}(x)$ and $(A^2)_{22}(x)$ on the phase $a$ is shown in panel e).}
    \label{fig:A}
\end{figure}

At this point let us briefly comment why $|\xi\rangle=\Omega_1 |g_1\rangle + \Omega_2 |g_2\rangle$ is a good choice for the vector that generates $|D_1(x)\rangle,|D_2(x)\rangle$ by Eqs.~\eqref{eq:genD1} and~\eqref{eq:genD2}. Let us for example consider  $|\xi\rangle = |g_3\rangle $. It leads to
\begin{align}
   |{D}_1(x)\rangle \sim & -\Omega_2 \sin(k_Lx+a)\ |g_1\rangle +  \Omega_1 \sin k_Lx\ |g_2\rangle  \label{eq:badD1} \\
   |{D}_2(x)\rangle \sim &  \Omega_1 \Omega_3 \sin k_Lx\ |g_1\rangle + \Omega_2 \Omega_3 \sin(k_Lx+a)\ |g_2\rangle\nonumber  \\ & - [\Omega_1^2 \sin^2 k_Lx + \Omega_2^2 \sin^2(k_Lx+a)]\ |g_3\rangle \label{eq:badD2}
   \end{align}
which is apparently analytically simpler than the previous results (\ref{eq:bestD1}-\ref{eq:bestD2}). In fact it gives a simple, analytic calculations of coefficients of $A(x)$ and $A^2(x)$ [see Appendix \ref{app:alpha}]. However, the limit  $a\to 0$ agrees with Eq.~\eqref{eq:darkD1indep} only up to a sign, namely both $\ket{g_1}$ and $\ket{g_2}$ components flip their signs when $k_L x = n\pi$, $n$ -- integer making them only piecewise constant and, in particular, discontinuous. For that reason, the derivatives in the definition of $A(x)$, Eq.~\eqref{eq:A} are ill-defined as $a\to 0$.

\subsection{Gauge potentials}
\label{subsec:gauge}

Having chosen the dark state basis  $|D_i(x)\rangle$'s (\ref{eq:bestD1}-\ref{eq:bestD2}), one finds the gauge potential $A(x)$ with Eq.~\eqref{eq:A}. For the dark-state projected Hamiltonian $H_2$ one restricts $A(x)$ and $A^2(x)$ to the upper left $2\times 2$ block. The coefficients $A_{ij}(x)$ for $i,j<3$ clearly do not depend on the choice of the bright states phase, as evident from \eqref{eq:A}.  The same holds for $(A^2)_{ij}(x)$ \cite{Ruseckas2005}:
\begin{align}
    (A^2)_{ij}(x)=&-\hbar^2 \sum_{M(x)\in{\cal{B}}(x)}\langle D_i | M'(x)\rangle \langle M | D_j'(x)\rangle =\nonumber \\
    =& \hbar^2 \langle D_i'(x) | D_j'(x) \rangle 
    \label{eq:proofA2}
\end{align}

Let us first consider a special case of $a=0$. With the position-independent $|D_1(x)\rangle$ and $|D_2(x)\rangle$ given by \eqref{eq:darkD2indep}, the coefficients of $2\times 2$ projections of $A(x)$ and $A^2(x)$ included in $\eqref{eq:H2}$ are all zero except for the $(A^2)_{22}(x)$ which is equal to:
\begin{equation}
    (A^2)_{22}(x)=\left(\frac{\epsilon\cos k_Lx}{\epsilon^2+\sin^2 k_Lx}\right)^2,
\label{eq:A22Vna}
\end{equation}
with $\epsilon={\Omega_3}/{\sqrt{\Omega_1^2+\Omega_2^2}}=\Omega_p/\Omega_c$, in analogy to $\Lambda$ system. 
The Hamiltonian $H_2$ is then a direct sum of two decoupled $D_1, D_2$ channels. The Hamiltonian~\eqref{eq:ham_dressed2} for the particle in the $D_1$ channel is that of a freely moving particle \begin{equation}
    H=-\frac{\hbar^2}{2m}\frac{d^2}{dx^2}.
\label{eq:A11lambda}
\end{equation}
with running waves eigenfunctions of the form $d_1(x)=\exp(iqx).$
The Hamiltonian $H_2$ for the particle in the $D_2$ channel reduces to the movement in the scalar potential given by $(A^2)_{22}(x)$:
\begin{equation}
    H=-\frac{\hbar^2}{2m}\frac{d^2}{dx^2}+(A^2)_{22}(x).
\label{eq:A22lambda}
\end{equation}
The above potential, given by \eqref{eq:A22Vna}, is precisely the subwavelength comb potential which appears for $\Lambda$ system construction~\cite{Lacki2016}.  It is shown in Fig.~\ref{fig:A}a).

For $a\neq 0$  the coefficients $A_{12}(x)$ and $(A^2)_{12}(x)$ are  non-zero, and the two channels $D_1$, $D_2$ are coupled. Fig.~\ref{fig:A}b) shows the spatial dependence of  $A(x)$ and $A^2(x)$ for a small $a=0.05$ value (for that $a$ value,  the $(A^2)_{22}(x)$ resembles that for $a=0$). The
potentials change for a larger value of $a$ as shown in  Fig.~\ref{fig:A} in panels (c) and (d) for  $a=0.435$.
The potentials $(A^{2})_{11}$ and $(A^{2})_{22}$ are in the form of a comb, with $(A^{2})_{22}$ being much larger than $(A^{2})_{11}$ (the remnant of vanishing   $(A^2)_{11}$ in the limiting case $a=0$). This is also evident from Fig.~\ref{fig:A}e) which shows the maximum height of $(A^{2})_{11}$ and $(A^{2})_{22}$ as the function of $a$.  
All the above potentials are clearly $\lambda_L/2$-periodic implying that the dark state-only model $H_2$ has period $\lambda_L/2$ -- half of the period of the full model $H$.

The potential shapes depend obviously on the choice of the basis in the dark subspace. In the Appendix this issue is discussed further.

\begin{figure}[t!]
    \centering
    \includegraphics[width=0.48\linewidth]{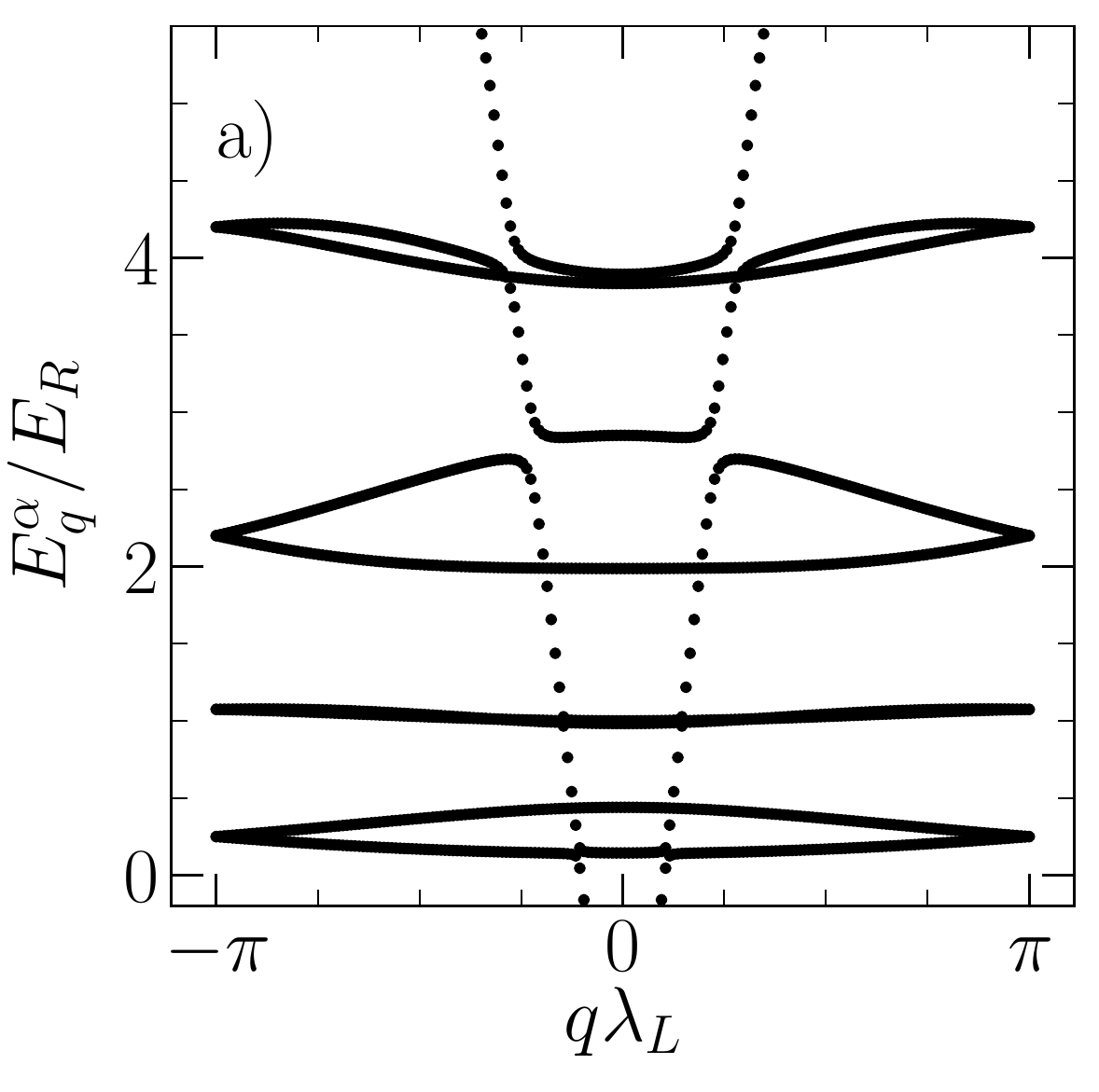} 
    \includegraphics[width=0.48\linewidth]{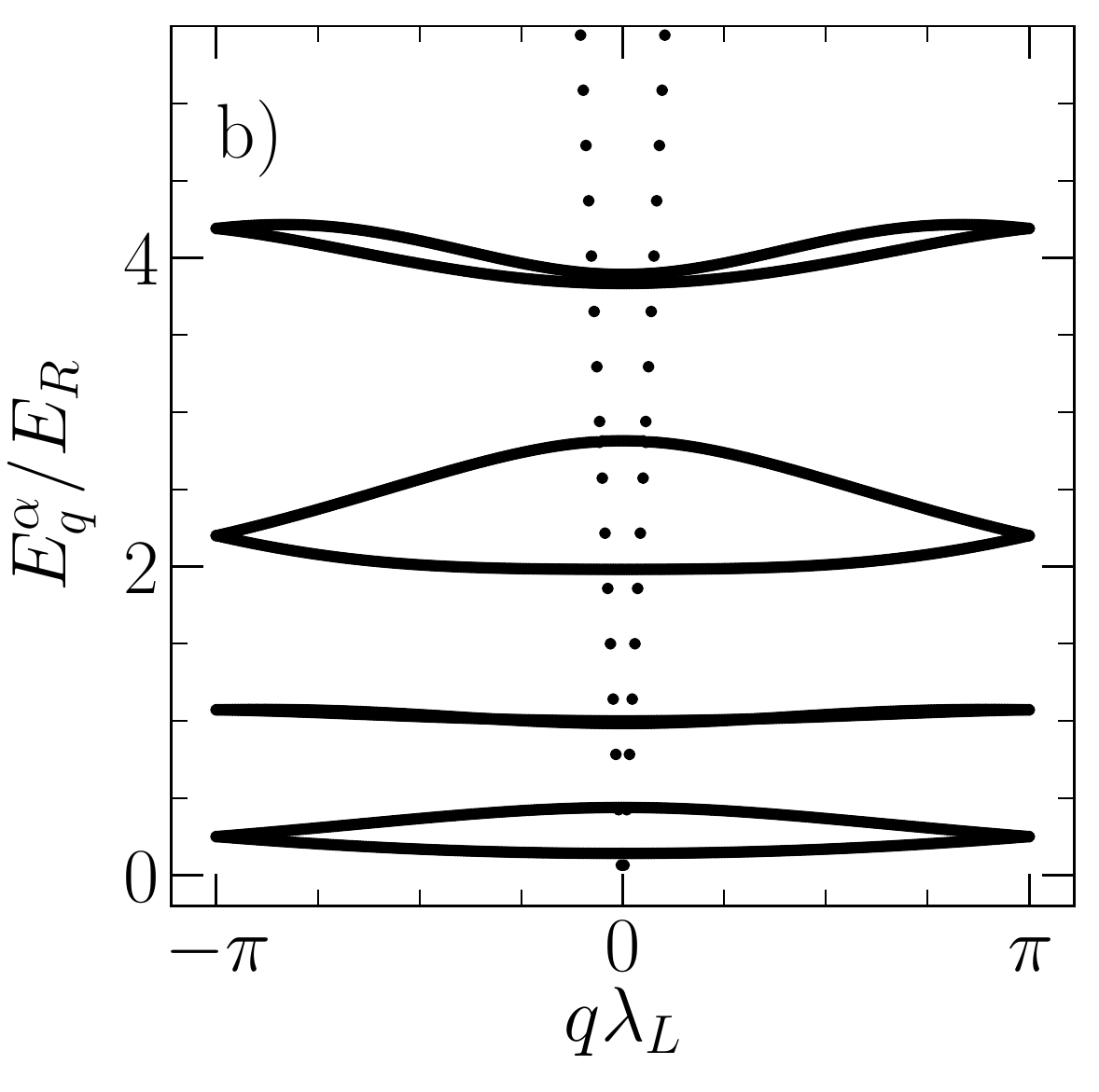} \\ 
    \includegraphics[width=0.48\linewidth]{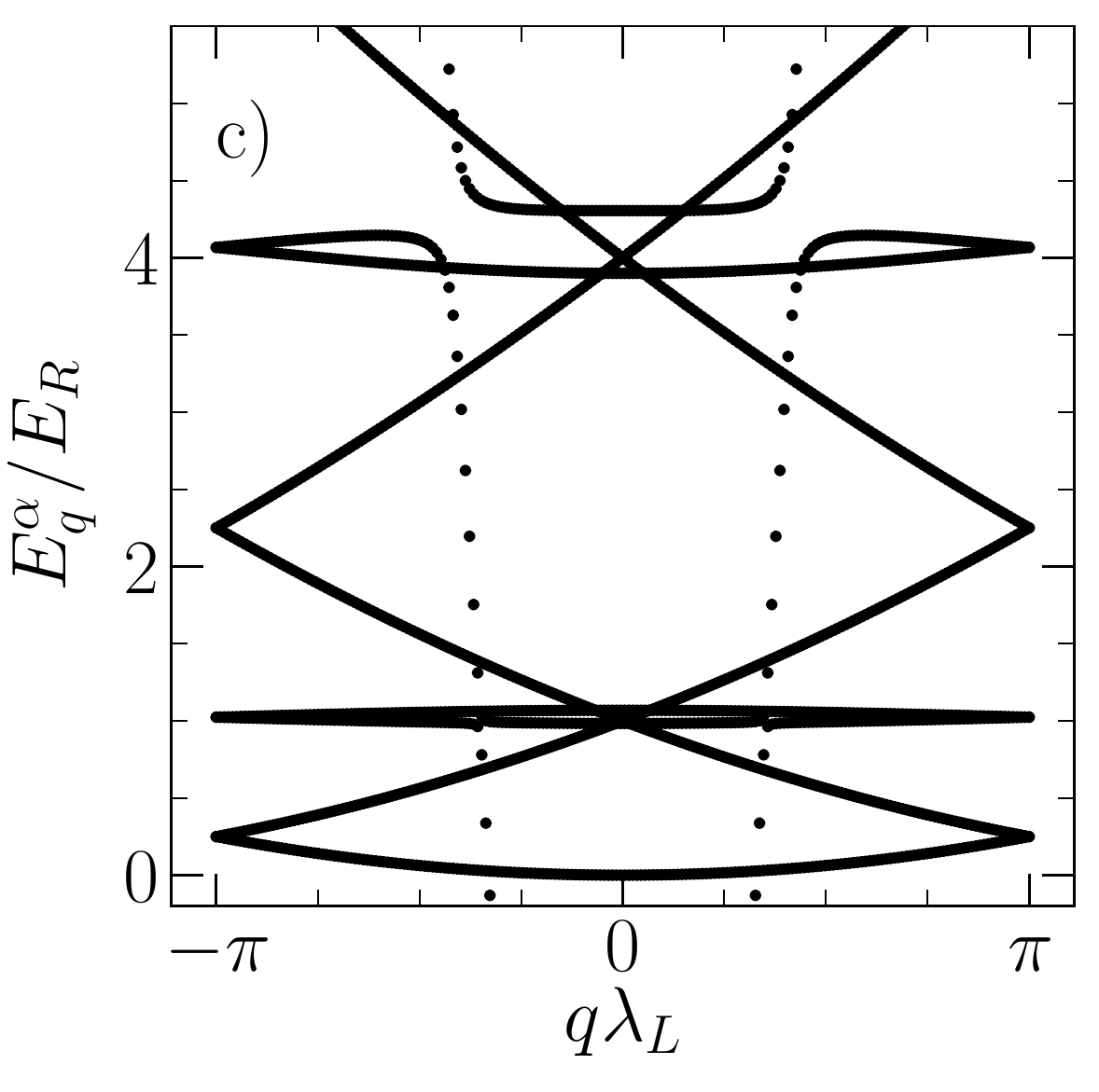} 
    \includegraphics[width=0.48\linewidth]{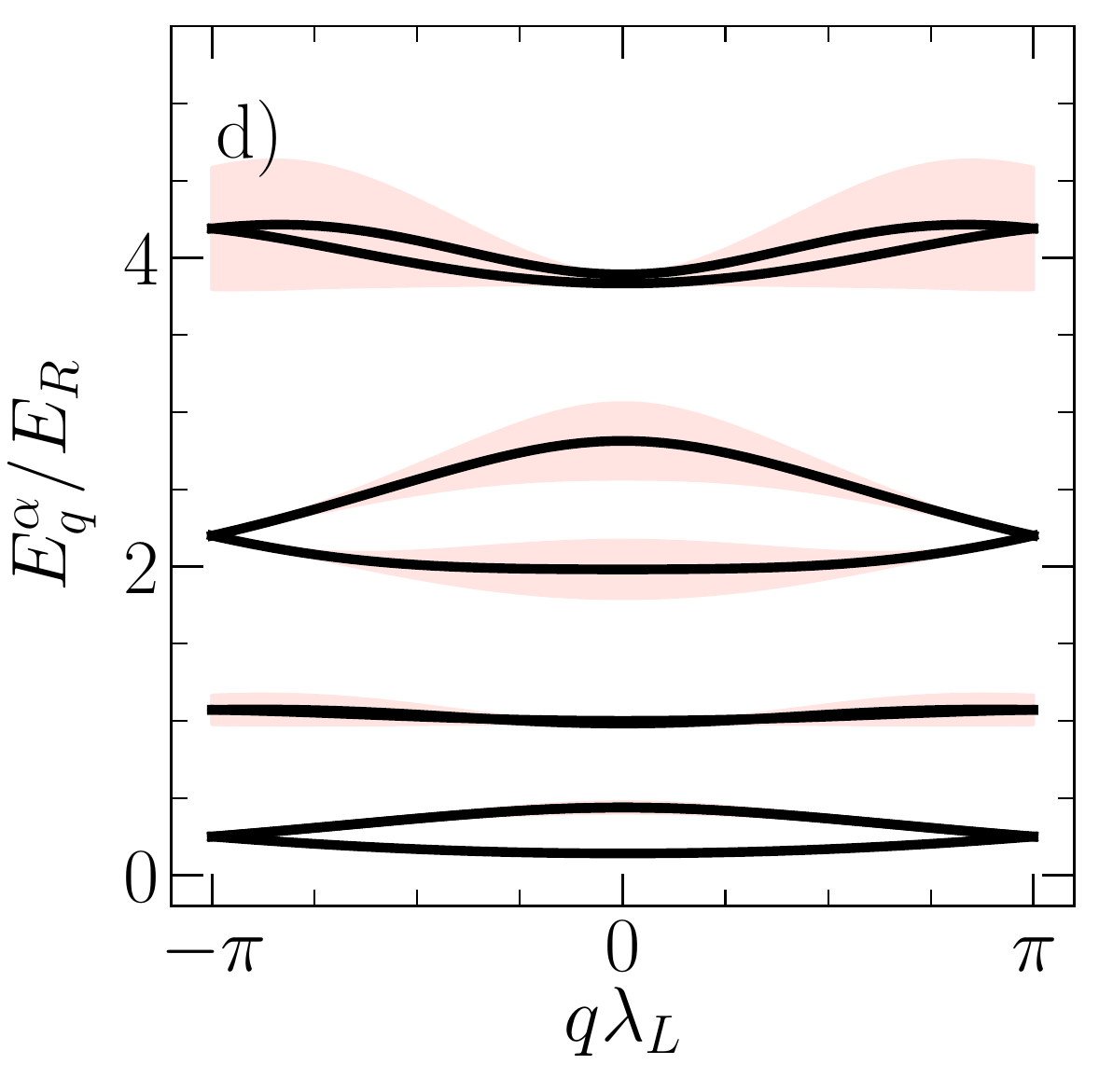}  
    
    \caption{Band structure of the model~\eqref{eq:ham_bare} for $\Omega_1:\Omega_2:\Omega_3=50:20:5$. Panel a) shows the spectrum for $\Omega_1=1000E_R$ and $a=0.435$. Panel b) $\Omega_1=5000E_R$ and $a=0.435$. Panel c) shows the dark-state only limit of the band structure for $\Omega_1=5000E_R$ and $a=0$. While in A0-c) $\Gamma_e=0$  in panel d) we show the band structure for $\Gamma_e=1000E_R$ and $\Omega_1=5000E_R$. The black line within bands shows the $\Re E_q^\alpha$ and the red region is given by the $\Re E_q^\alpha \pm 50\Im E_q^\alpha$ curves, denoting $q$-dependent losses due to spontaneous emission. For clarity the bright state eigenvalues have been removed from this panel.
     }
    \label{fig:bandStructures}
\end{figure}

\section{The band structure}
\label{sec:bandStructure}

\subsection{General considerations}

We discuss the band structure of the full model~\eqref{eq:ham_bare}. In numerical calculations we work directly with the full Hamiltonian~\eqref{eq:ham_bare}. The BO decomposition into dark and bright states, and in particular Hamiltonian $H_2$ in Eq.~\eqref{eq:H2} is instrumental for the interpretation of the results.

We look for the quasiperiodic Bloch eigenstates of the  $\lambda_L$-periodic model~\eqref{eq:ham_bare} directly in $|g_i\rangle,|e\rangle$ basis:

\begin{align}
B_q(x)=&e^{iqx}\left(\sum_{i=1}^3 b_{g_i}(x) |g_i\rangle + b_e(x)|e\rangle\right),\\
\equiv & e^{iqx}[b_{g_1}(x),b_{g_2}(x),b_{g_3}(x),b_{e}(x)]^T. 
\label{eq:Bq}
\end{align}
The period of the Hamiltonian and the Bloch theory guarantee that $b_*(x)$ are $\lambda_L$-periodic functions and $q$ is the quasimomentum $q\in BZ_1=[-\pi/\lambda_L,\pi/\lambda_L)$, where $BZ_1$ is the Brillouin zone.
Looking at Eq.~\eqref{eq:ham_bare} and spatial dependence of $\Omega_i$'s one finds an extra parity symmetry:  coefficients $b_{g_i}, b_e$ are actually all $\lambda_L/2$-periodic or  $\lambda_L/2$-antiperiodic, and if $b_{g_1},b_{g_2}$ are $\lambda_L/2$-periodic then $b_{e},b_{g_3}$ are $\lambda_L/2$-antiperiodic and vice versa.

For the numerical formulation of the eigenproblem for Hamiltonian \eqref{eq:ham_bare} the Fourier series expansion of $b_*(x)$ is used. It puts the Hamiltonian, $H$, in a sparse matrix form, which is diagonalized using  standard numerical packages \footnote{In this work we have used standard scipy diagonalization function \textsc{eigs}.}. The different eigenvalues $E_q^\alpha$ are indexed by $\alpha$ for each value of $q\in BZ_1$. 

Cosider first $\Gamma_e=0$ case -- the corresponding  band structure obtained for some particular values of $\Omega_i$ is depicted in Fig.~\ref{fig:bandStructures}.
When  $||\vb{\Omega}||$ dominates other energy scales, the full spectrum contains energy levels that can be traced back to:  $E_-$ bright states,  $E_+$ bright states and dark states $D_1,D_2$, effectively described by Eq.~\eqref{eq:H2}.

Fig.~\ref{fig:bandStructures}a) shows the section of the band structure at low energies above zero that feature a series of bands and two nearly vertical lines of eigenvalues intersecting them. We identify the bands with the dark subspace, as they can be reproduced with identical Bloch bands computation for dark-state-projected model $H_2$. The vertical lines originate from $E_-$ bright states and are modelled by the Hamiltonian:
\begin{equation}\label{eq:bareBm}
    H_{B-}=-\frac{\hbar^2}{2m_a}\frac{d^2}{dx^2}+E_-(x).
\end{equation}
Energy levels that can be traced to $E_+(x)\gg 0 $ are absent in the figure. This channel does not have any energy levels at energies close to 0, however, its eigenstates can be coupled to dark states off-resonantly.

The dark subspace bands are two-valued. This comes from the fact that coefficients of $A(x),A^2$ as in \eqref{eq:ham_dressed2} are of period $\lambda_L/2$ and the Bloch theory applied to Fig.~\ref{fig:bandStructures} assumes twice larger lattice period of $\lambda_L$, natural period of Eq.~\eqref{eq:ham_bare}. 
Complementary results from $H_2$ model could be obtained from Bloch theory with lattice period $\lambda_L/2$ and a larger Brillouin zone $BZ_2=[-2\pi/\lambda_L,2\pi/\lambda_L]$. 
This would yield dark only Bloch eigenfunctions of the form:
\begin{align}\label{eq:Blochdark}
    (B_D)_q^\alpha =&e^{iqx}\left( b_{D_1}(x)|D_1(x)\rangle+b_{D_2}(x)|D_2(x)\rangle\right)\nonumber\\
    \equiv & e^{iqx}[b_{D_1}(x),b_{D_2}(x)]^T
\end{align}
where $q\in BZ_2$ and $b_*$ are $\lambda_L/2$-periodic. 
We note that the above vector, when re-expressed in the $|g_1\rangle,|g_2\rangle,|g_3\rangle,|e\rangle$ is only $\lambda_L$-periodic just as $|D_1(x)\rangle, |D_2(x)\rangle$.
The 4-channel computation for large $||\mathbf{\Omega}||$ yields good approximation of the above. For quasimomenta $q,q'\in BZ_2, q\in BZ_1$ such that $|q-q'|=2\pi/\lambda_L$ the $\lambda_L$-periodic Bloch theory treatment of full $H$ ascribes them both to a single $q\in BZ_1$. Such folding has already appeared for a special case of a $\Lambda$ system (see~\cite{Lacki2016}) and is not unique to the tripod configuration, it is a simple consequence of the mismatch between the  dark-state lattice constant and the period of the model.

Couplings due to $A(x)$ between dark states and the resonant $E_-$ states lead to small avoided crossings [clearly visible in Fig.~\ref{fig:bandStructures}a)]. Fig.~\ref{fig:bandStructures}b) illustrates the fact that for larger $\Omega_i$ the avoided crossings with the $E_-$ bright state get narrower indicating an even better isolation of the dark subspace. 
The improving separation between the dark subspace and the
$E_-$ bright band with increasing  $\Omega_i$
 is easily understood from the model~\eqref{eq:H2}. The avoided crossings appear between low lying dark-states and the highly excited $E_-$  with the same $q\in BZ_1$. Increasing $||\mathbf{\Omega}||$ pushes the $E_-$ manifold towards more negative energies. As a result for larger $||\mathbf{\Omega}||$ the wave vector describing freely moving $E_-$ with approximately zero energy is more and more oscillating. This reduces the coupling to dark state Bloch vector via $A(x)$.

For a comparison we show also bands corresponding to $a=0$ case in Fig.~\ref{fig:bandStructures}c). 
Here the dark state $|D_1\rangle$ is position independent [see~\eqref{eq:darkD1indep}], the spectrum is a sum of that of a freely moving particle in channel $D_1$ [see Fig.~\ref{fig:bandStructures}c)] and a $D_2$-particle feeling the presence of the potential~\eqref{eq:A22Vna}. The two spectra intersect each other with no avoided crossings forming between them. The bright state line also does not couple to position-independent $D_1$ through $A(x)$ \eqref{eq:A}. The avoided crossing prominent in Fig.~\ref{fig:bandStructures}c) is between $B_-$ and $D_2$ channel bands.

\subsection{Dark-bands lifetime}
\label{subsec:bands:lifetime}

Let us consider now the lifetime of different bands. When $\Gamma_e\neq 0,$ in the Hamiltonian \eqref{eq:ham_at_bare},
 the energies $E_\pm(x)$ of the bright state  channels~\eqref{eq:bright_e} that appear in~\eqref{eq:ham_dressed}, \eqref{eq:ham_dressed2} are complex, in particular  $\Im E_\pm(x) = -\Gamma_e/4$ for $\Delta=0$.  The diagonalization of the model~\eqref{eq:ham_at_bare} focusing on low-lying dark state band reveals strong $q$-dependence of $\Im E_q^\alpha$ [as shown in  Fig.~\ref{fig:bandStructures}d)]. Here we assume $\Gamma_e=1000E_R$ and red shaded areas  $\Im E_q^\alpha$ are multiplied by 50 to make them more visible.
 The $q$-dependence for the $\Lambda$ system was already described in~\cite{Lacki2016} and indirectly observed experimentally~\cite{Wang2018}. 
 
\begin{figure}[htb]
    \centering
    \includegraphics[width=0.48\linewidth]{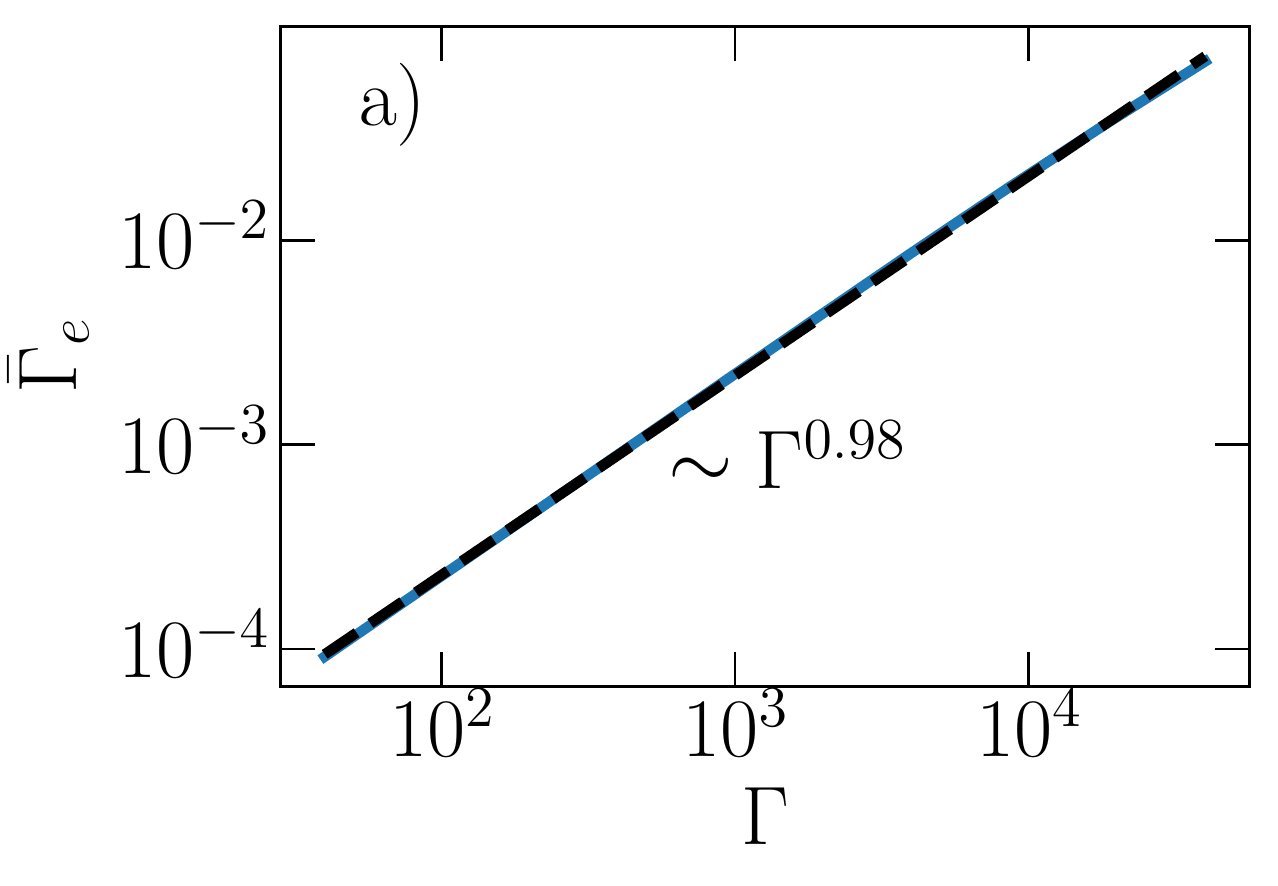}
    \includegraphics[width=0.48\linewidth]{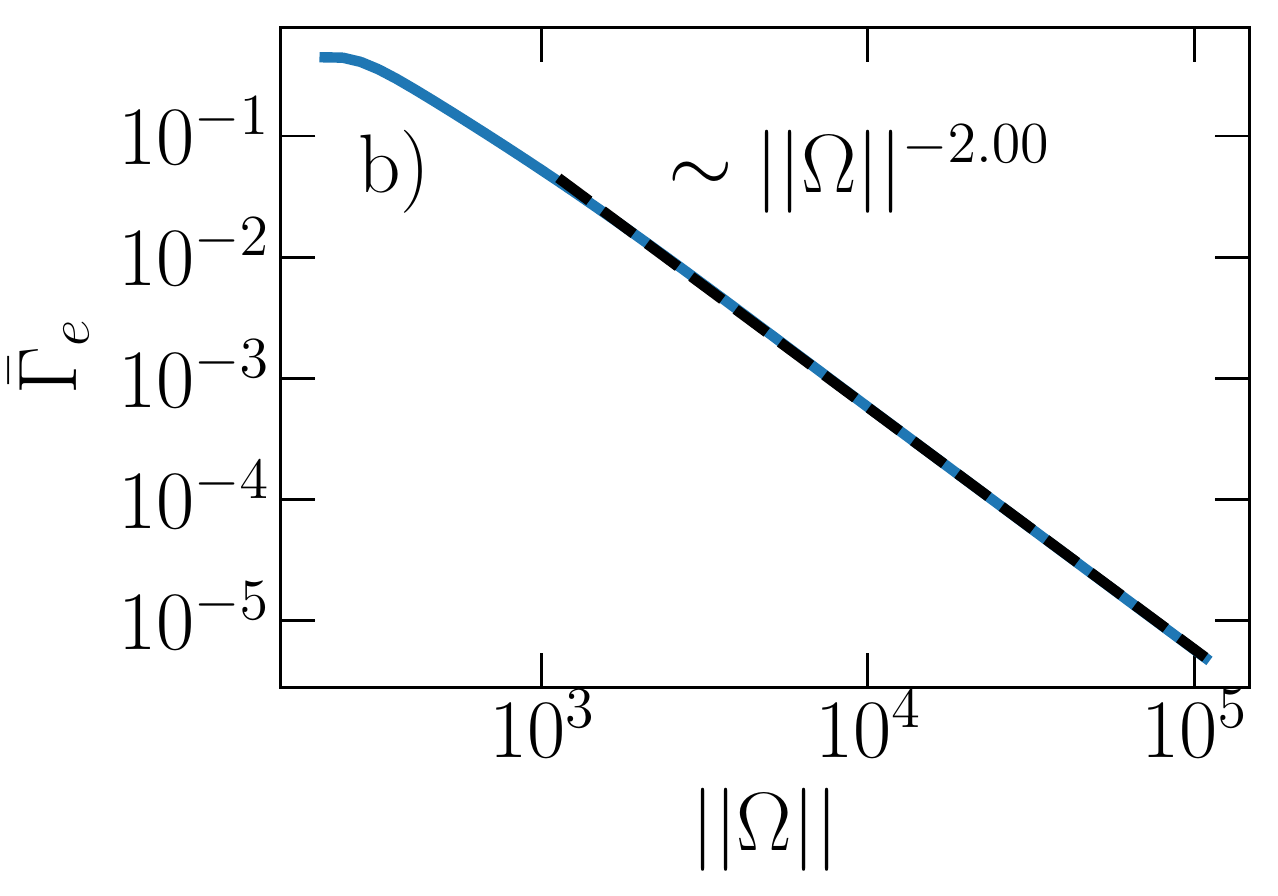}\\
    \includegraphics[width=0.98\linewidth]{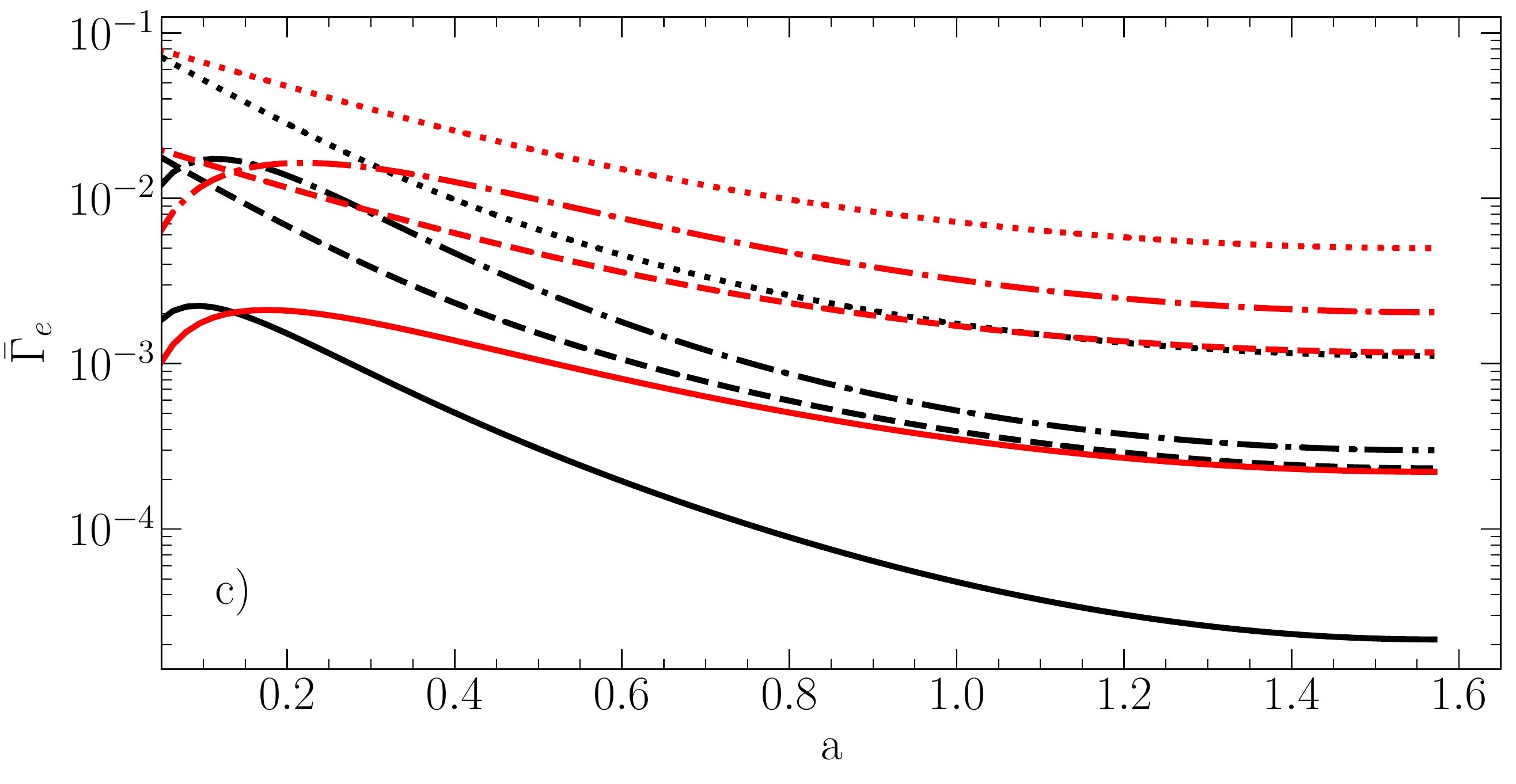}
    \caption{Panel a)  The $q$ averaged decay rate  $\bar{\Gamma}_e$  for $(\Omega_1=5000E_R, \Omega_2=2000E_R, \Omega_3=500 E_R$) as the function of $\Gamma_e$ for $2^\text{nd}$ band. Panel b) shows the same as the function of $||\vb{\Omega}||$ for same relative ratio of $\Omega_i$'s and $\Gamma_e=1000E_R$. The panel c) shows $\bar{\Gamma}_e $ as the function of $a$ for $\Gamma_e=1000 E_R$ and two sets of $\Omega_i = (5000E_R, \Omega_2, 500 E_R)$, for four first bands. Black (darker) lines correspond to $\Omega_2 = 2000 E_R$, while red (lighter) lines to $\Omega_2 = 1000 E_R$. The bands are depicted subsequently as solid, dashed, dashed-dotted and dotted lines.}
    \label{fig:gamma}
\end{figure}
The total lifetime of the gas populating a certain band is approximated by the inverse of the $q$-averaged decay rate
\begin{equation}
\bar{\Gamma}_e = -\frac{2}{\text{vol}(BZ_1)}\int_{q\in BZ_1}   \Im E_q^\alpha \textrm{d}q.
\end{equation}
 Figure~\ref{fig:gamma}a) shows the numerically obtained dependence of $\bar{\Gamma}_e$  on $\Gamma_e$ for $\Omega_1=5000E_R, \Omega_2=2000E_R, \Omega_3=500 E_R$. Panel b) shows the dependence on $||\vb{\Omega}||$ assuming a fixed ratio $\Omega_1:\Omega_2:\Omega_3=50:20:5$.
Just like in the $\Lambda$ system~\cite{Lacki2016}, one observes that $\bar{\Gamma}_e\propto \Gamma_e$ and $||\vb{\Omega}||^{-2}.$  

A good approximation for $\Im E_q^\alpha$ follows from the second-order perturbation theory arguments. The imaginary contribution to the energy of the dark state Bloch function $(B_D)_q^\alpha(x)$ is
\begin{equation}
\Im \Delta E_q^\alpha=-\sum_\beta\sum_{\sigma\in\pm} \frac{\Gamma_e}{2}\frac{|\langle (B_D)_q^\alpha|H_c| E_{q,\sigma}^\beta  \rangle|^2}{(E_q^\alpha - \Re (E_\sigma)_q^\beta)^2+\frac{\Gamma_e^2}{4}}
\label{eq:2nd}
\end{equation}
where $E_{\sigma,q}^\beta$ refer to bright state eigenvectors in potentials $E_\sigma(x), \sigma\in \pm$ with the same quasimomentum $q$. $H_c$ in \eqref{eq:2nd} contains all the non-diagonal terms in Hamiltonian~\eqref{eq:H2}. 
For the vast majority of states indexed by $\beta$ the $\Gamma_e^2$ term in the denominator may be neglected. Moreover, the sum is dominated by 
by bright states $E_-(x)$ with energy close to $\max E_-(x) \propto ||\mathbf{\Omega}||$, and bright states of $E_+(x)$ with energy close to $\min E_+(x)\propto ||\mathbf{\Omega}||$. This qualitatively explains the observed dependence on $\Gamma_e$ and $||\mathbf{\Omega}||.$

The coupling in the numerator of~\eqref{eq:2nd} depends on  $A(x)$ terms in~\eqref{eq:ham_dressed2}. It is greatly increased if the coefficients of $A$ responsible for the coupling of the dark state to the bright states are sharply peaked and large as it happens in the $a\rightarrow 0$ limit leading then to
 larger losses 
  as shown in Fig.~\ref{fig:gamma}c). Already $a=0.435$ offers order of magnitude longer lifetime than $\Lambda$ system case, $a=0$. 
We also note that the ratio of $\Omega_1:\Omega_2:\Omega_3$ strongly affects the expected lifetime, particularly for large $a$. Fig.~\ref{fig:gamma}c) presents the simulated $\bar{\Gamma}_e$ for first four bands for $\vb{\Omega}=(5000E_R,2000E_R,500E_R)$ (black lines) and $\vb{\Omega}=(5000E_R,1000E_R,500E_R)$ (red lines). Reducing the ratio $\Omega_2/\Omega_3$ from 4 to 2 results in the order of magnitude shorter lifetime for large $a\approx \pi/2$.

\subsection{Spin decomposition of Bloch bands}
\label{subsec:bands:spinstructure}

\begin{figure}[htbp]
    \centering
    \includegraphics[width=0.48\linewidth]{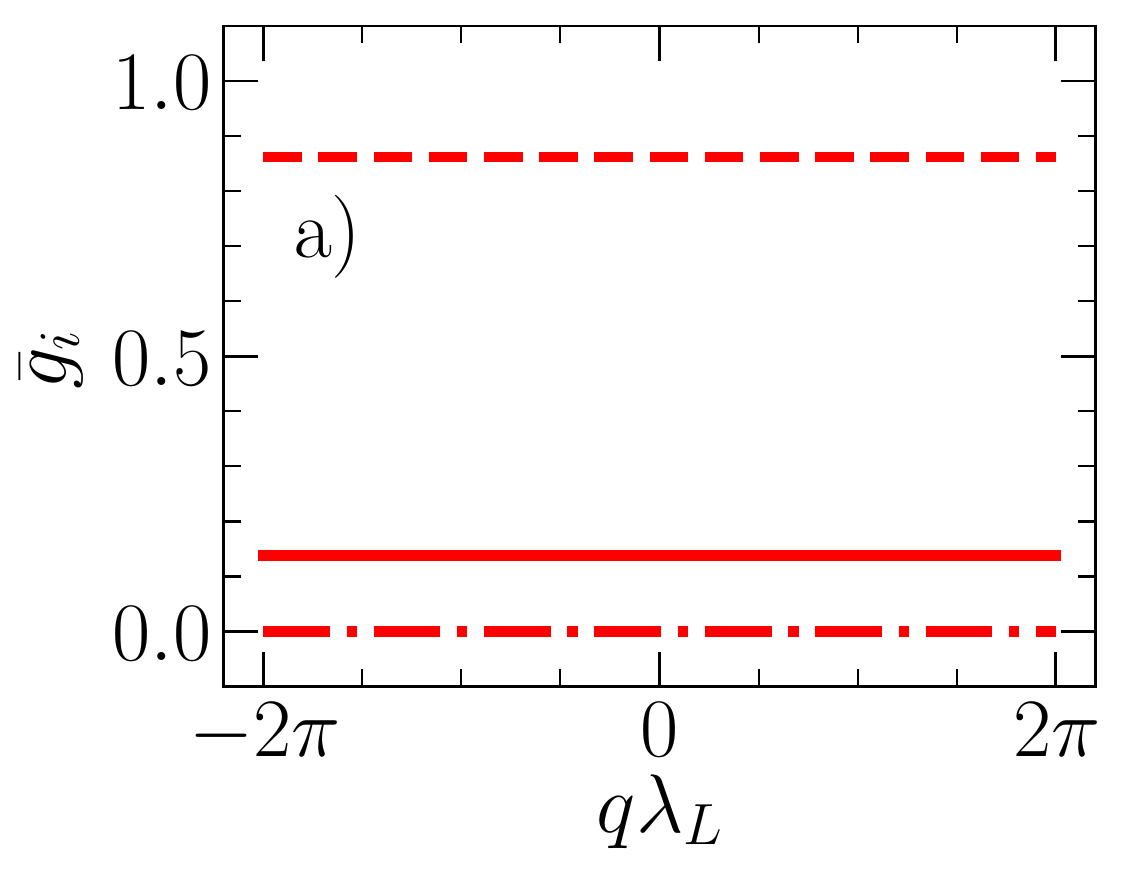} 
    \includegraphics[width=0.48\linewidth]{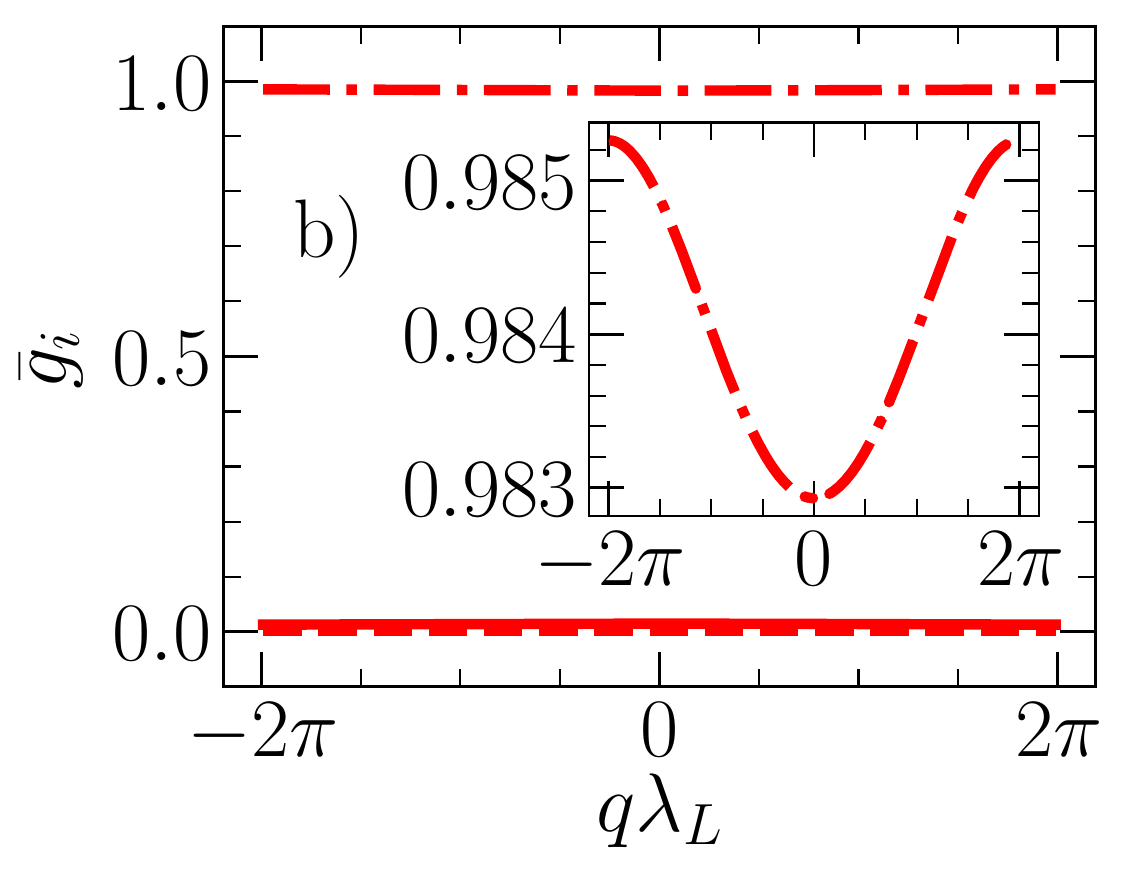} \\ 
    \includegraphics[width=0.48\linewidth]{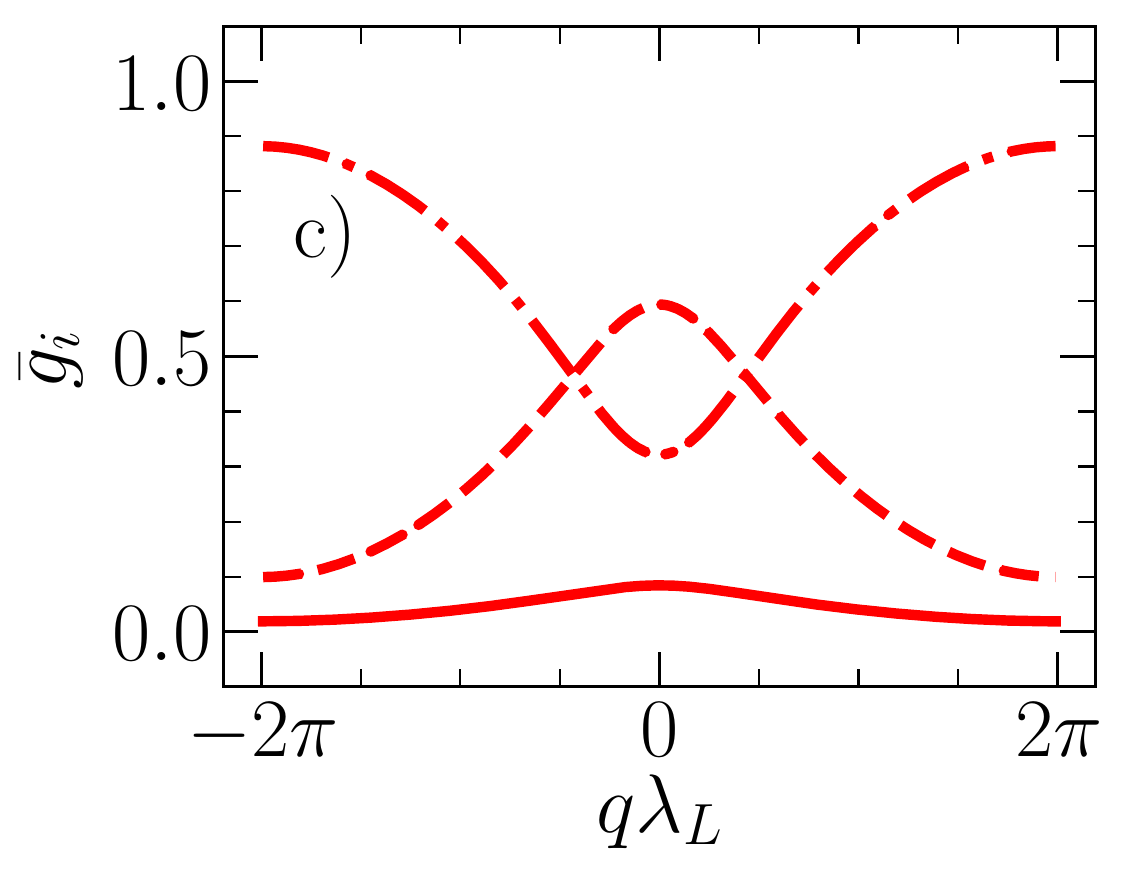} 
    \includegraphics[width=0.48\linewidth]{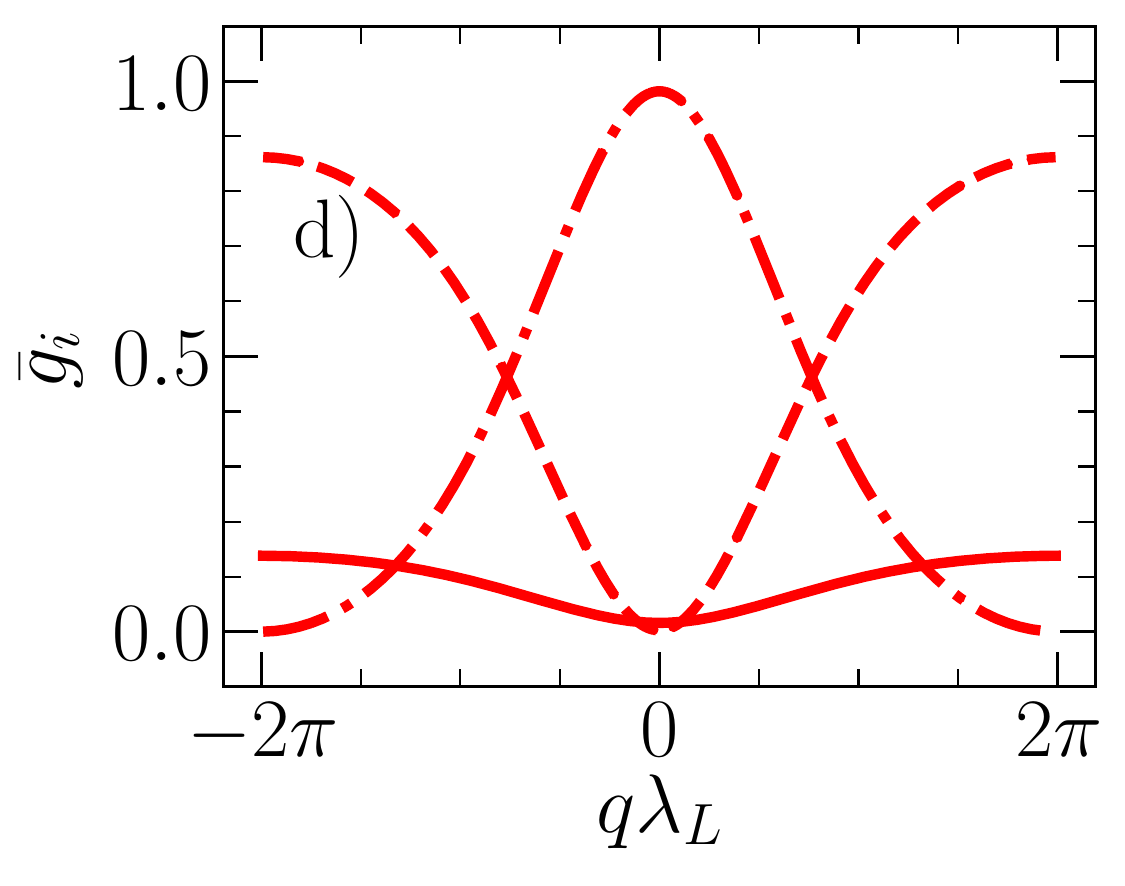}  
    
    \caption{Decomposition of Bloch states in $BZ_2$ into $\ket{g_i}$, where solid, dashed and dot-dashed lines correspond to, respectively, $i=1, 2, 3$. Top row a), b) corresponds to $a=0$, the bottom row c), d) $a=0.435$. Left column a), c) shows lowest bands, while the right column b), d) -- first excited bands. {The inset in b) presents a weak dependence of $\ket{g_i}$ on this panel using $\ket{g_3}$ as an example.}
    }
    \label{fig:gi}
\end{figure}

Let us discuss the decomposition of Bloch eigenvectors into atomic states $|g_1\rangle,|g_2\rangle,|g_3\rangle,|e\rangle$.  We use again our exemplary set of parameters  $\vb{\Omega}=(5000E_R,2000E_R,500E_R)$ for an illustration, taking also $\Gamma_e=0$. Figure~\ref{fig:gi} shows the averages $\bar{g}_i=\int_0^{\lambda_L} |\langle B_q^\alpha|g_i\rangle|^2$ for different quasimomenta within first two bands of $a=0$ and $a=0.435$ systems. 

For $a=0$ the lowest ``band'' is actually a portion of the parabolic energy dependence of a freely moving particle. It forms a closed band as soon as $a\neq 0$. The $\bar{g}_i$ are in that case constant and given by the constant coefficients of Eq.~\eqref{eq:darkD1indep}. For the first excited band, that is the lowest band in the $D_2$ channel, the dependence on $q$ is very small [panel b)].
When $a\neq 0$ is increased towards the final value $a=0.435$ the well defined bands are formed, as in Fig.~\ref{fig:bandStructures}a). Different parts of each band intersect other energy levels, experiencing avoided crossings in different ways. For example the $D_1$ and $D_2$ channels at energy close to $1E_R$  for $q=0$ in Fig.~\ref{fig:bandStructures}c) experience transition through the avoided crossing before well separated bands in Fig.~\ref{fig:bandStructures}a) are formed. At the same time energy levels at $q=\pm \pi/\lambda_L\in BZ_1$ remain nearly unaffected by other energy levels. This is the reason for the observed strong dependence on $q$ of decompositions $\bar{g}_i$ in Fig.~\ref{fig:gi} c) and d) for $a=0.435$.
Specifically, one can observe that decompositions into $\bar{g}_i$  of the first excited band for $q=0 \in BZ_2$ resembles the decomposition of the $D_2$ channel in Fig.~\ref{fig:gi}b), and for $q=\pm 2\pi/\lambda_L\in BZ_2$ that of $D_1$ in Fig.~\ref{fig:gi}a).  At the same time the lowest band shows strong dependence of $\bar{g}_i$ on $q$ which however does not approximate $D_1$ or $D_2$ for any $q$.

The variation of the overlap of Bloch vectors on $|g_1\rangle, |g_2\rangle, |g_3\rangle$ within the same band is directly observable. Consider a bosonic, non-interacting gas cooled down to the least energetic state of the band. Denote the quasimomentum of such a state by $q_0$. If an extra potential $H_{\textrm{tilt}}=-Fx$ is added (see \cite{Dahan1996}) to the Hamiltonian~\eqref{eq:ham_bare} then a steady drift of the quasimomentum $q(t) = q_0+Ft/\hbar $ occurs, allowing one to reach the desired value of $q$ by controlling the application time of $H_{\textrm{tilt}}$.  The value of $F\lambda_L$ should be much smaller than the energy gap to other bands, to prevent populating them.  The spin decomposition can be studied by turning off the lasers responsible for Rabi frequencies $\Omega_i$ and splitting the atomic cloud in $|g_1\rangle, |g_2\rangle, |g_3\rangle $ components by a magnetic field gradient. 

The response of the tripod system to gradient $H_{\textrm{tilt}}$ would form a coherent, time-dependent transfer of populations of atomic states $|g_1\rangle, |g_2\rangle ,|g_3\rangle $, a feature which is easily measurable.

\section{The tight binding model}
\label{subsec:bands:tightBinding}
A tight-binding model conveniently describes movement of the particles populating a particular band. We describe first the  construction of the Wannier functions in analogy to the textbook Wannier function calculation for a  cosine-squared optical lattice~\cite{Kohn1959, Kivelson1982, Marzari2012}. 

\subsection{{The construction of Wannier functions in two-dimensional dark subspace}}

We start with the basis of all the dark-state Bloch functions $\{(B_D)_{q}^{\alpha}:  {q\in BZ_2}\}$ for a particular band $\alpha$ of $H_2$. Then the Wannier function can be expressed as:

\begin{equation}
W_{n}^{\alpha}(x)=N\int_{q\in BZ_2}(B_D)_{q}^{\alpha}(x)e^{i\theta_{q,n}}\textrm{d}q,
\label{eq:Wny}
\end{equation}
the index $n$ denotes localization over $n$-th lattice site, $x_n=x_0+n\lambda_L/2,$ and $N$ ensures that $\int_\mathbb{R} |W_n^\alpha(x)|^2\textrm{d}x =1$. The functional dependence of phases $\theta_{q,n}$ on $q$ has to be chosen to localize the $W_n^\alpha$. To find it, we adapt the method by Kivelson~\cite{Kivelson1982}. The $H_2$ Hamiltonian is considered under periodic boundary conditions in a box of a sufficient total length $L$. This discretizes the Brillouin Zone $BZ_2 \to [0,2\pi/L,\ldots, 4\pi/\lambda_L ) $.  We construct the $L\times L$ matrix
\begin{equation} \label{eq:kivelson_matrix}
M_{q,q'}=\langle (B_D)_q^\alpha | e^{2\pi i x/L} | (B_D)_{q'}^\alpha\rangle,\quad  q,q'\in \textrm{BZ}_2.  
\end{equation}
Its eigenvalues are complex phases of the type $\exp[(2\pi i x_n)/L]$ -- that determines $x_0$. The corresponding eigenvector then defines the values of $\theta_{q,n}$ that localize $W_{n}^{\alpha}$ around the location $x_n$.  We have verified that the obtained Wannier functions are exponentially localized around $x_n$, as expected for this procedure~\cite{Kivelson1982}.

It is also worth noting that using this method Wannier functions can be computed directly within $\lambda_L$-periodic Bloch theory ($q\in BZ_1$) by including both branches of a folded band while computing \eqref{eq:kivelson_matrix} matrix elements.

For a single-channel problem with a periodic potential, one can calculate a single Wannier function e.g. $W_0(x)$ and then use a discrete translation $W_n(x):=W_0(x-(x_n - x_0))$   to complete the basis. For the tripod system this is also a possibility but a special care should be taken when applying translation to $W_0^\alpha(x)$ given by Eq.~\eqref{eq:Wny} when expressed in $|g_i\rangle,|e\rangle$ basis. Indeed if we expand $W_n^\alpha(x)$ in $|D_i(x)\rangle$ basis:
\begin{equation}
W_0^\alpha(x) = w_1(x) |D_1(x)\rangle + w_2(x) |D_2(x)\rangle 
\label{eq:W0s}
\end{equation}
then one can shift the $w_i(x)$ functions only and resum the $W_n(x)$ to obtain
\begin{equation}
W_n^\alpha(x) = w_1(x-n\lambda_L/2) |D_1(x)\rangle + w_2(x-n\lambda_L/2) |D_2(x)\rangle .
\label{eq:shiftWgood}
\end{equation}
Instead one might attempt to translate 
the entire $W_0^\alpha(x)$ obtaining
\begin{align}
w_1(x-&n\lambda_L/2) |D_1(x-n\lambda_L/2)\rangle +\nonumber \\ &+w_2(x-n\lambda_L/2) |D_2(x-n\lambda_L/2)\rangle \neq W_n^\alpha(x).
\label{eq:shiftWbad}
\end{align}
As $|D_i(x)\rangle$ are only $\lambda_L$-periodic, both approaches are not equivalent for odd $n$.  The former approach leading to \eqref{eq:shiftWgood} is a proper one, as it 
corresponds to a shift of the coefficients of the Wannier function by the Hamiltonian $H_2$ lattice constant for a single-valued band defined by $H_2$ with BZ of $BZ_2$. We have verified that for odd $n$, $W_n^\alpha$ translated  as in \eqref{eq:shiftWbad} is not orthogonal to $W_0^\alpha(x)$.

The properties of Wannier functions are discussed further in Section \ref{sec:wanprop} , after constructing the tight binding Hamiltonian description.

\begin{figure}[htbp]
    \centering
    \includegraphics[width=0.95\linewidth]{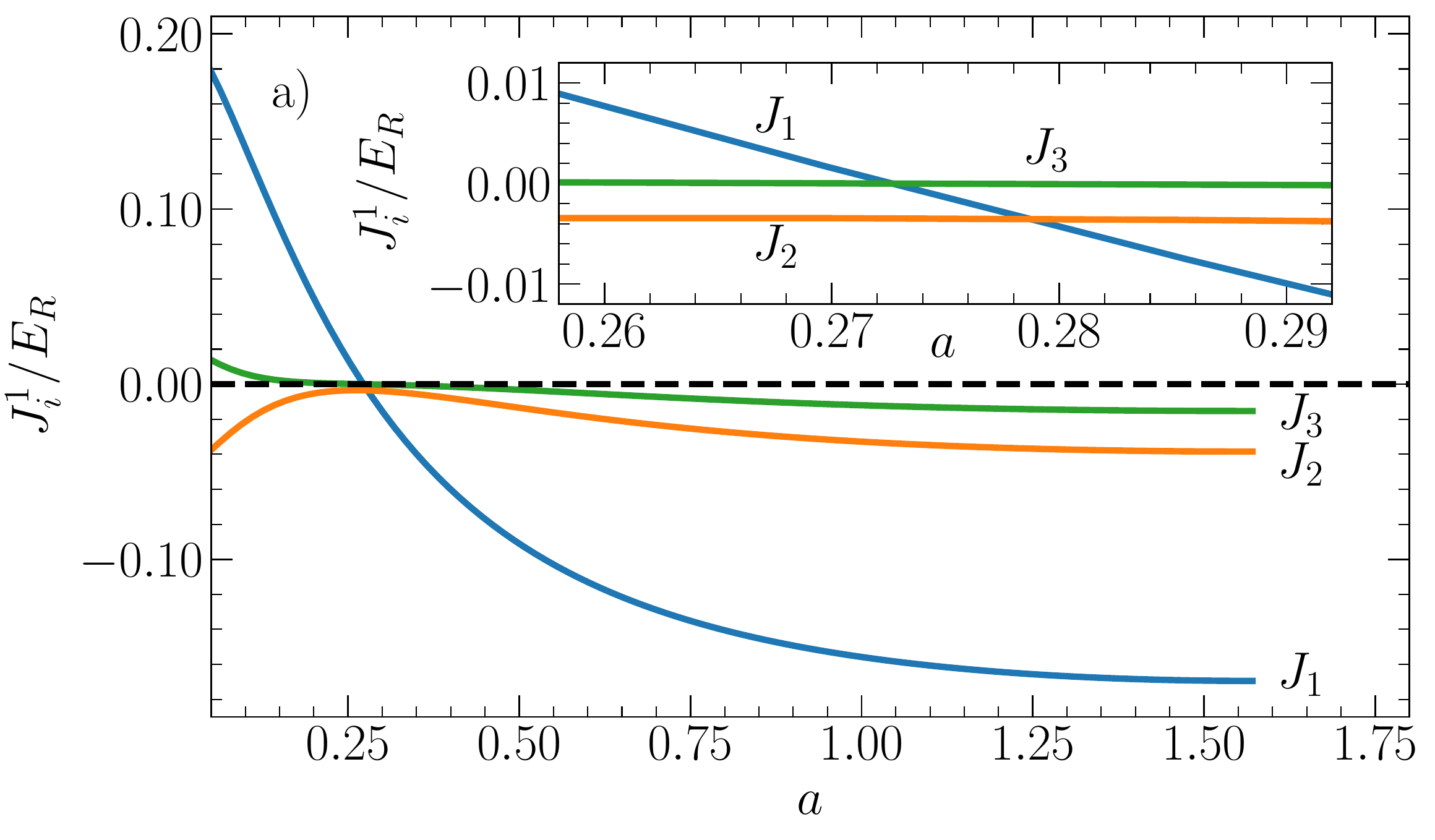} \\
    \includegraphics[width=0.95\linewidth]{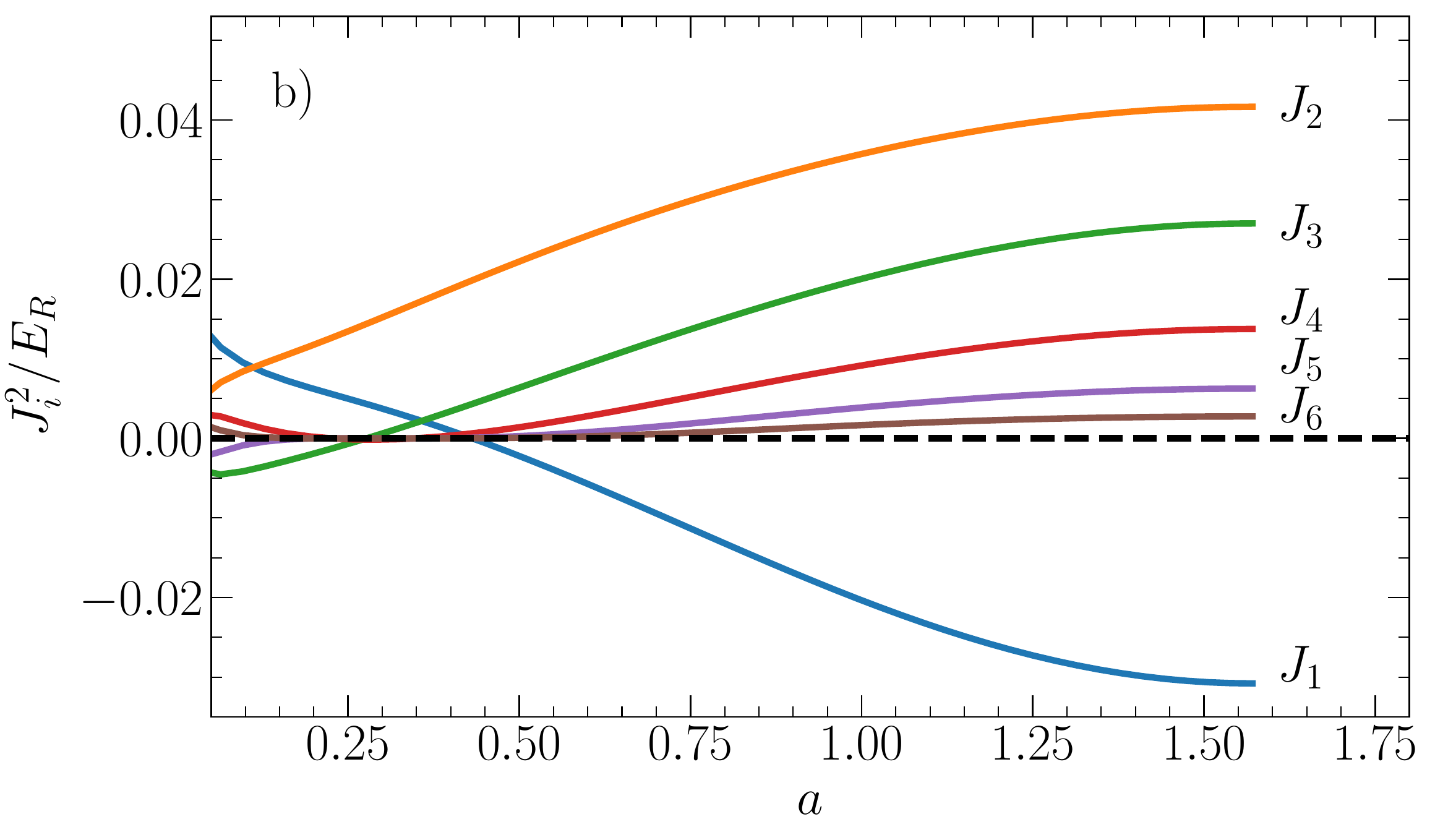} \\
    \includegraphics[width=0.95\linewidth]{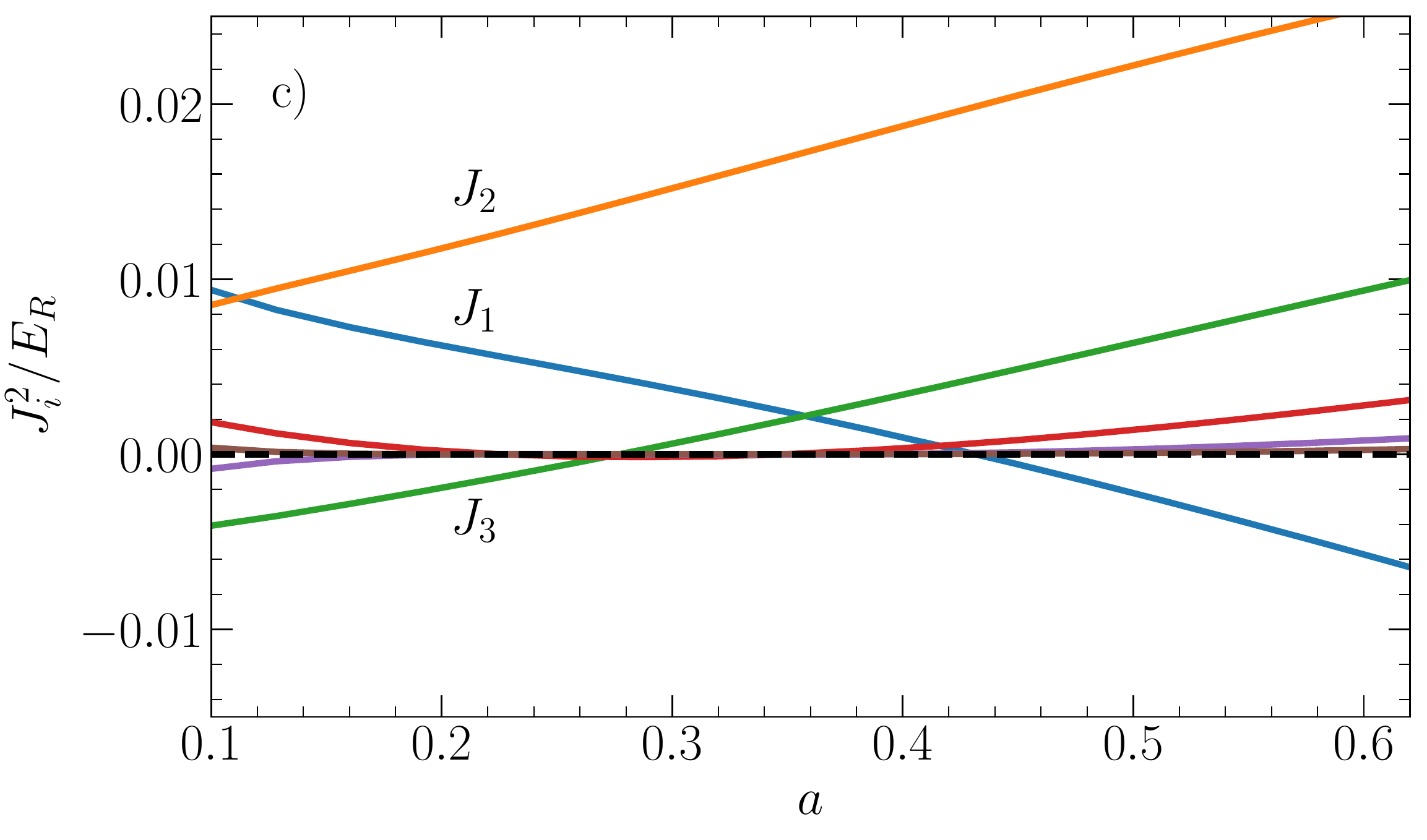} \\
    \includegraphics[width=0.95\linewidth]{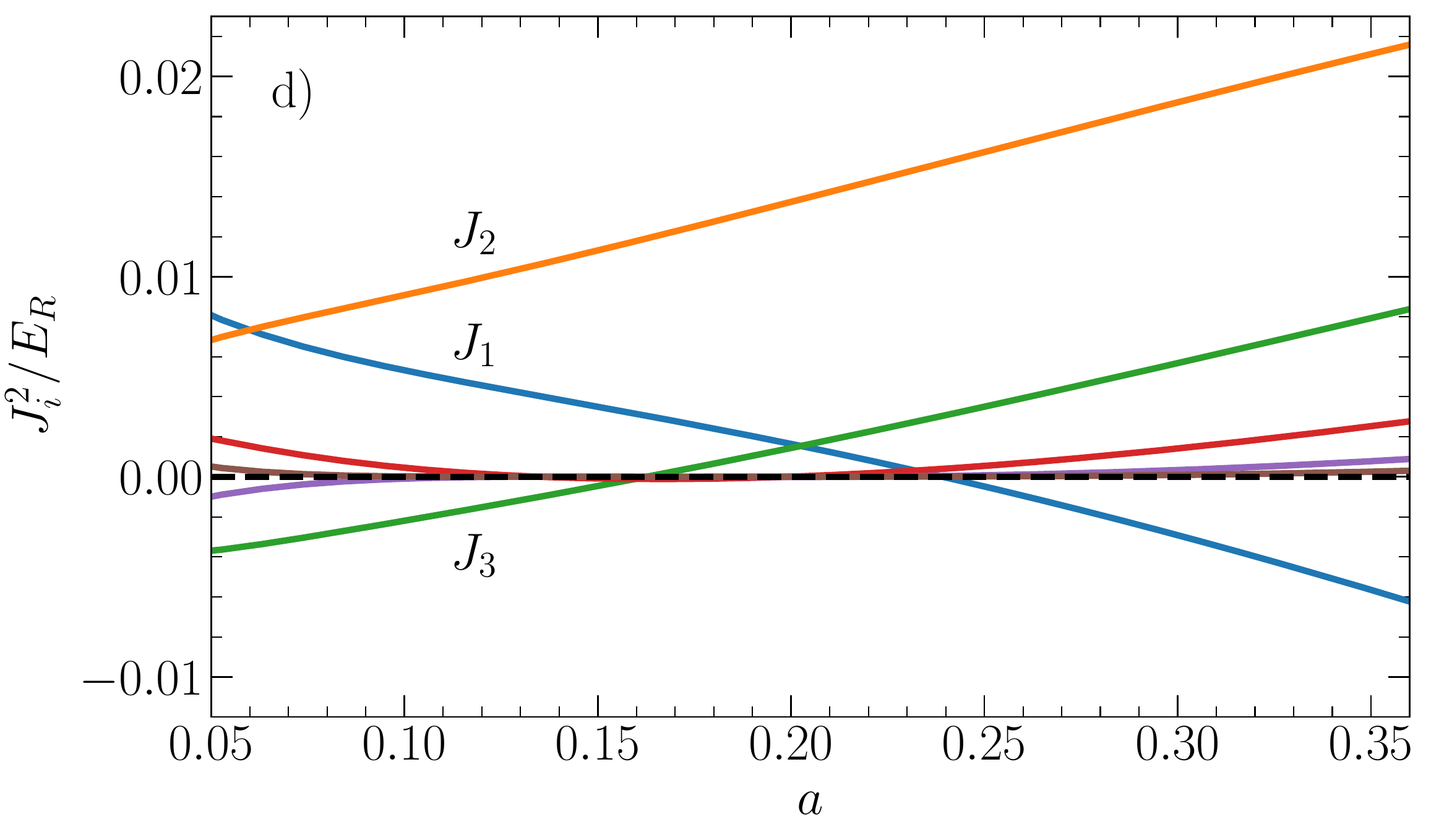}
    \caption{Different hopping amplitudes $J^\alpha_i$ as a function of $a$ -- the relative phase difference between $\Omega_1(x)$ and $\Omega_2(x)$. Panel a) shows the ground band $\alpha=1$ for $\Omega_1:\Omega_2:\Omega_3=50:20:5$. Panel b) presents the same systems, but in the first excited band $\alpha=2$. The magnification of this panel with a narrower range of $a$ is shown in panel c). Panel d) shows the analogous range of $a$, but for $\Omega_1:\Omega_2:\Omega_3=50:40:5$.}
    \label{fig:jty}
\end{figure}

\subsection{{Hopping amplitudes}}
\label{hophop}

The Hamiltonian $H_2$, restricted to band $\alpha$ when expressed in the basis $W_{n}^{\alpha}(x)$ transforms to

\begin{equation}
H_{\textrm{hopp},\alpha}	=	-\sum\limits _{n,m}J_{nm}^{\alpha\alpha}(\hat{a}_{n}^{\alpha})^{\dagger}\hat{a}_{m}^{\alpha}+H.c.\label{eq:BHsp}
\end{equation}
where:
\begin{align} J_{nm}^{\alpha\alpha}	&=	-\int dx\ (W_{n}^{\alpha})^\dagger(x) H_2(x) W_{m}^{\alpha}(x)=\nonumber \\
&=-\frac{1}{\textrm{vol}(BZ_2)}	\int_{BZ_2}e^{i(m-n)q\lambda_L/2}E^{\alpha}_q\textrm{d}q.
\label{ha}
\end{align}
The second equality is true  assuming that the global phase factors of $W_n^\alpha(x)$ are defined by \eqref{eq:shiftWgood}. The hopping {amplitudes} depend only on the distance between sites $n,m$ {so one can simplify the notation by defining:}
\begin{equation}
    J_{|n-m|}^\alpha := J_{n,m}^{\alpha\alpha}.
\end{equation}

The hopping amplitudes (referred simply as hoppings later on) may be directly calculated
from their definition \eqref{ha} using previously determined Wannier functions. Their exponential tail requires, however, a special care for accurate determination of $J^\alpha_i$, important, in particular, for $i>1$. However,
 the hoppings $J_i^\alpha$ can be read from the band energies, for $q\in BZ_2$:
\begin{equation}\label{eq:Jq}
E_q^\alpha=E_0-2J_1^\alpha\cos(q \lambda_L/2)-2J_2^\alpha\cos(2q \lambda_L/2)-\ldots
\end{equation}

Calculation directly in $\ket{g_i},\ket{e}$ basis followed by band unfolding suffices to determine $J_i^\alpha$ as well.  As the $J_i$-s are defined with respect to $q\in BZ_2$, a mistake in the unfolding of $q\in BZ_1$ would lead to a sign flip of $J_i$-s with an odd $i$. The Bloch vectors obtained in the 4-channel calculation can be projected back onto the $|D_1(x)\rangle,|D_2(x)\rangle$ space. The quasi-periodicity of the coefficients  
\begin{equation}
b_{D_i}(x+\lambda_L/2)=\exp(iq\lambda_L/2)b_{D_i}(x)
\label{eq:quasiper}
\end{equation}
allows to distinguish the two Bloch states $q,q'\in BZ_2$, $|q-q'|=2\pi/\lambda_L$ that correspont to the same point in $BZ_1$.

Unfortunately, the band unfolding by assigning of $q',q\in BZ_2$ is gauge-dependent.  Applying Eq.~\eqref{eq:quasiper} uses a particular gauge during projection on $|D_i(x)\rangle$.
As a result when two eigenvectors for a particular $q\in BZ_1$ are being relabelled by $q,q'\in BZ_2$, $|q-q'|=2\pi/\lambda_L$ the assignment of $q,q'$ is reverse for $D_i$ as in Eqs.~\eqref{eq:bestD1},\eqref{eq:bestD2} and $D_i$ as in 
Eqs.~\eqref{eq:badD1},\eqref{eq:badD2}. This means that the dependence on $q$ of the quasienergy $E_q^\alpha$ present in Eq.~\eqref{eq:Jq} differs by a translation by $2\pi/\lambda_L$ and a sign flip in $J_n$'s: 
$J_n^\alpha  \to J_n^\alpha (-1)^n$. 
The ambiguity of signs of $J_i^\alpha$ is not in conflict with definition of $J_i^\alpha$ by means of the Wannier functions, Eq.~\eqref{ha}.  It is fully recovered when the $J_n^\alpha$ are computed from Eq.~\eqref{ha} in both gauges. 

We follow the gauge choice given  by Eqs.~(\eqref{eq:bestD1}-\eqref{eq:bestD2}) and calculate the tunnelings for low lying (and long living) bands. Again we discuss similar parameter values as before, i.e. $\vb{\Omega}=(5000,2000,500)E_R$.
Consider first the lowest band taking a familiar from the standard Bose-Hubbard model form -- compare Fig.~\ref{fig:bandStructures}. Not surprisingly $|J_2 |\ll J_1$ for most of values of the phase shift parameter $a$ thus nearest neighbor hopping dominates.
Interestingly, however, $J_1$ changes sign when $a$ is varied -- compare Fig.~\ref{fig:jty}a) which allows for realization of frustration as discussed in Section \ref{sec:frust}.    

In Fig.~\ref{fig:jty}b) and c)  we show the values of $J_i^\alpha$'s as the function of $a$ for 
the first excited, almost flat, band (compare Fig.~\ref{fig:bandStructures}) that results in  an unusual relation between $J_1$ and longer distance hopping amplitudes.  For $a<0.2$ or $a>0.45$, for this band, amplitudes for long-distance  hopping $J_i$ with $i>3$ are non-negligible indicating that the tight binding approach may be not a best choice in such a case. However for $a\in [0.2, 0.45]$ only $J_1,J_2,J_3$ can be considered for an accurate tight-binding model.
The next nearest hopping  $J_2$ is larger then nearest-neighbor amplitude $J_1$ and larger than  next-next-next nearest neighbor amplitude $J_3$. Only for $a\approx 0.2$ do $J_1$ and $J_2$ become comparable but are of opposite signs. 
Around $a=0.435$ a special situation occurs as $J_1\approx 0$. This is in agreement with band structures in Fig.~\ref{fig:bandStructures}, where the second band seems to be  "single valued" at the scale of the figure.

The Fig.~\ref{fig:jty}d) shows a similar calculation of hopping {amplitudes} for the excited band, but for the ratio of $\Omega_1:\Omega_2:\Omega_3=50:40:5$. The same configurations of amplitudes $J_i$ occur in a different range of the phase shift $a$ parameter. The corresponding interval for $a$ is $[0.1,0.22]$. Its location and size depends approximately linearly on the ratio $\Omega_3/\Omega_2$, as long as $\Omega_2<\Omega_1$ and $\Omega_2/\Omega_3\gg 1$. When $\Omega_2/\Omega_3 \approx 1$ the region of interest does not exist (naive interpolation puts it for $a>\pi/2$).
When $\Omega_2>\Omega_1$ the role of $\Omega_1,\Omega_2$ is reversed. 

\subsection{Frustration}
\label{sec:frust}

It is well known that a 1D spin-1/2 chain with $J_1$ (nn) and $J_2$ (nnn) interactions may exhibit frustration \cite{Mikeska} -- a situation in which it is not easy
to satisfy energetical minimalization of all the possible bonds \cite{Diep}. Such a chain maps into a nonpartite triangular ladder in which for negative tunnelings $J_1,J_2$ kinetic frustration occurs (\cite{Eckardt17} reviews the physics of periodically driven systems that enable a change of sign of the tunneling matrix elements, see also \cite{Sacha12}). In our situation the sign of $J_1$ (or any $J_i$ for odd $i$) can be inverted by a gauge transformation (reverting the sign of the every second Wannier function). It is thus more interesting that for the lowest band $J_2$ becomes negative
(antiferromagnetic). For most $a$ values $J_1$ dominates making frustration difficult to observe. However, since $J_1$ changes sign [around  $a=0.275$ for the chosen values of $\Omega_i$, see Fig.~\ref{fig:jty}a)] it becomes small and comparable to $J_2$ for nearby $a$ values leading to quite standard
frustrating system. Note that a change of sign of $J_1$ in triangular lattices was realized via periodic lattice shaking (see \cite{Eckardt17,Sacha12} and references therein) -- here no additional shaking is needed and frustrating conditions are are realised by changing the phase mismatch $a$.

Situation is equally interesting for the first excited band. Here,  compare
Fig.~\ref{fig:jty}, both the  nn, $J_1$ and nnnn, $J_3$ may change sign depending on $a$ value while $J_2$ remains positive and large.
Consider first the simplest situation when we adjust $a$ such that $J_3$ vanishes. The system maps to a triangular ladder with $J_2$ positive and regardless of the sign of $J_1$ no frustration occurs. This is again a manifestation of the fact that  the change of sign of every second Wannier functions is just a gauge transformation that changes the sign of $J_{2i+1}$ leaving the physics unaltered.

\begin{figure}[htb]
    \centering
    \includegraphics[width=0.99\linewidth]{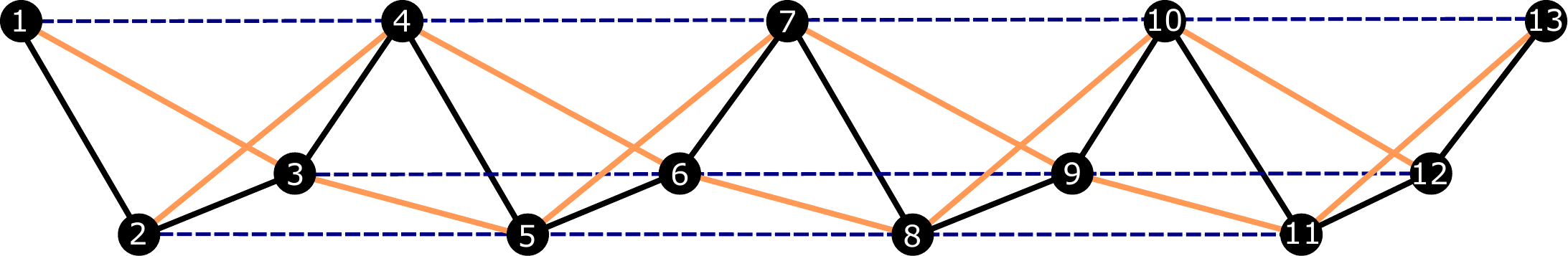}

    \caption{A mapping of a 1D chain with nn, nnn, and nnnn tunnelings into a tetrahedral linear compound with nearest neighbor tunnelings only. Along dashed blue lines $J_3$ tunnelings occur, orange lines correspond to $J_2$ while black connections yield the nearest neighbor terms $J_1$.
    }
    \label{fig:frustration}
\end{figure}

In the presence of $J_3$ the models becomes less obvious. The mapping on the triangular ladder does not work anymore. Instead one can map a 1D chain into a three dimensional tetrahedral linear compound as depicted in Fig.~\ref{fig:frustration}. Such a representation allows us for a better visualization of a competition between different hopping terms. Now it is easy to see that if the signs of $J_1$ and $J_3$ are different the system will frustrate as one cannot minimize energetically the $i,i+1$ and $i,i+3$ bonds. On the other hand, in the interval of $a$ values where $J_1$ and $J_3$ are of the same sign, no kinetic frustration occurs.

\subsection{{The properties of Wannier functions}}
\label{sec:wanprop}

\begin{figure}[htb]
    \centering
    \includegraphics[width=0.48\linewidth]{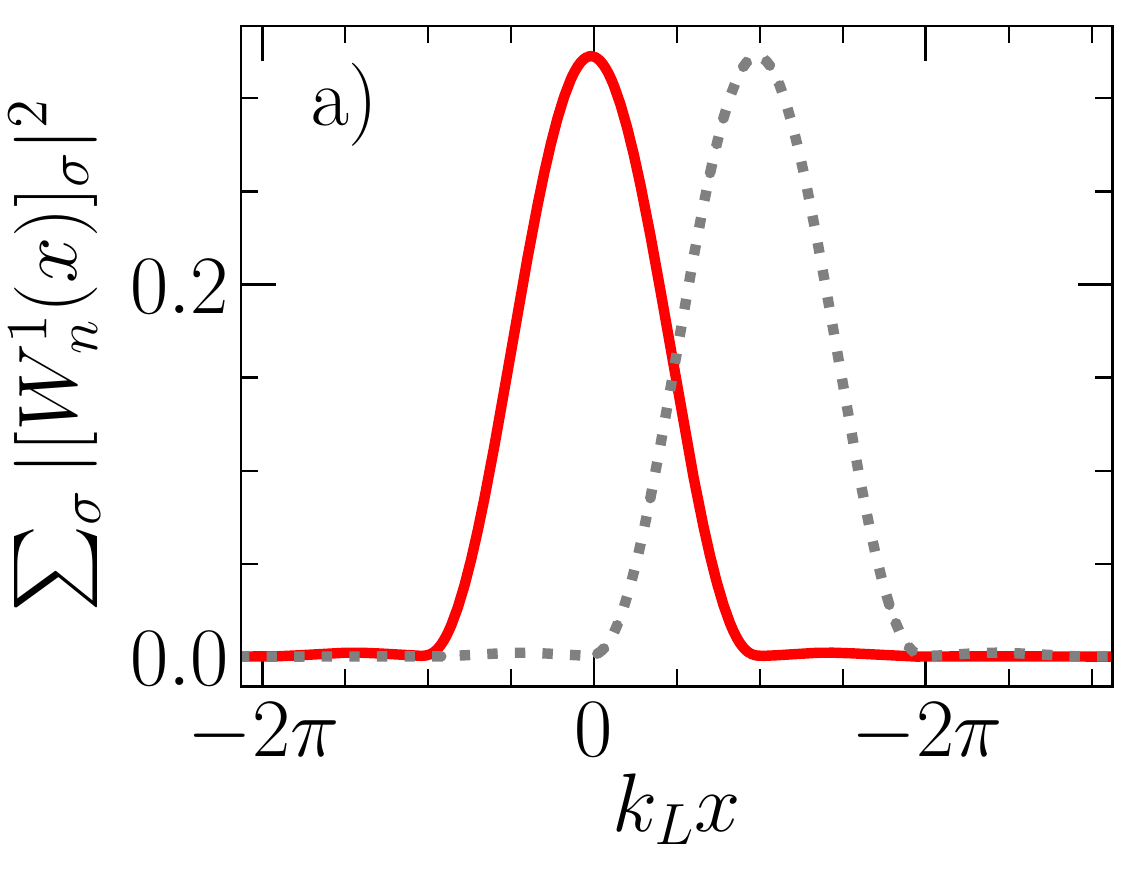}
    \includegraphics[width=0.48\linewidth]{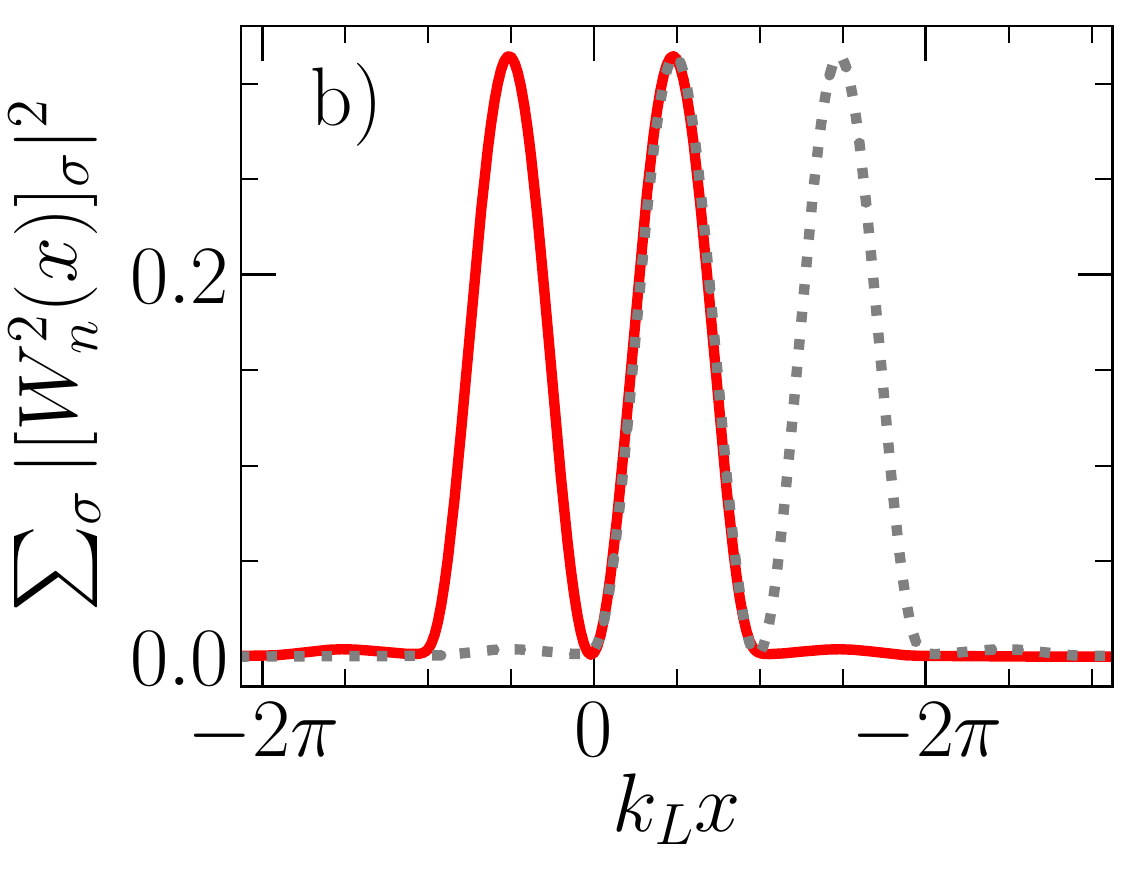} \\
    \includegraphics[width=0.48\linewidth]{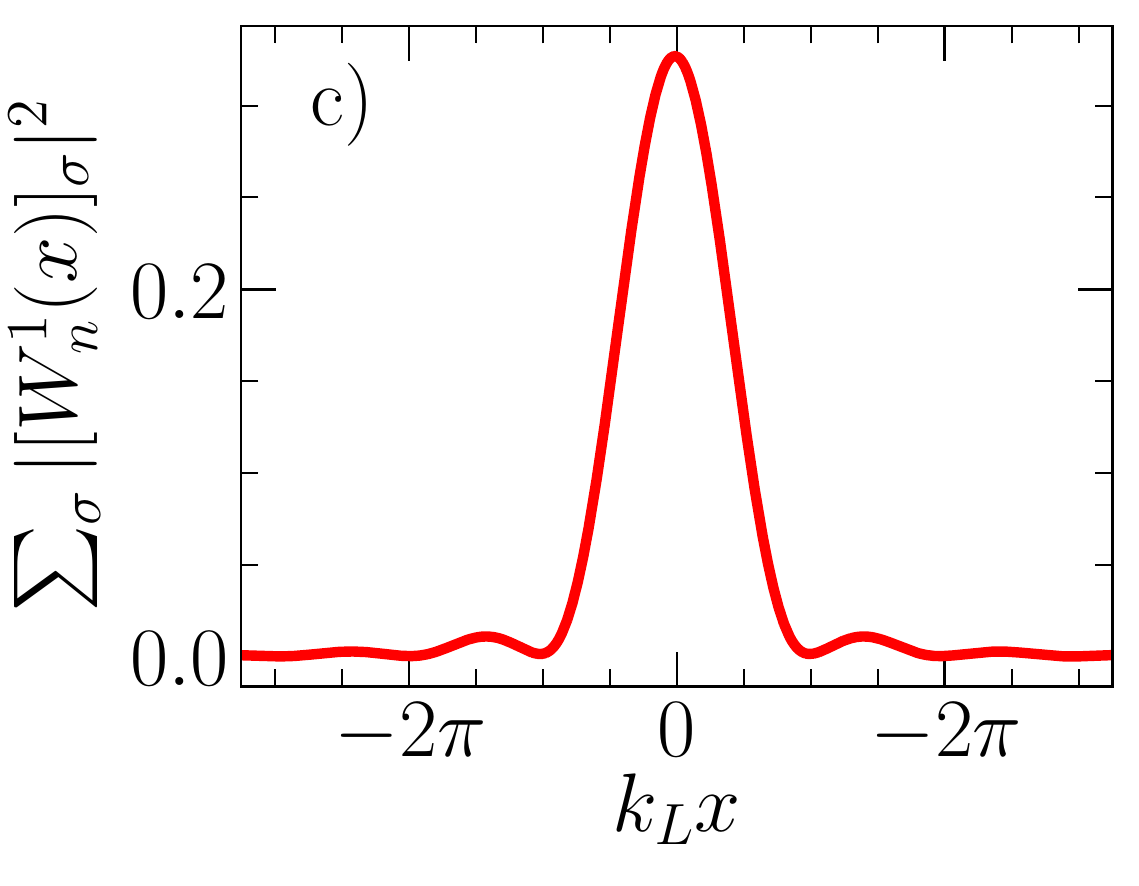}
    \includegraphics[width=0.48\linewidth]{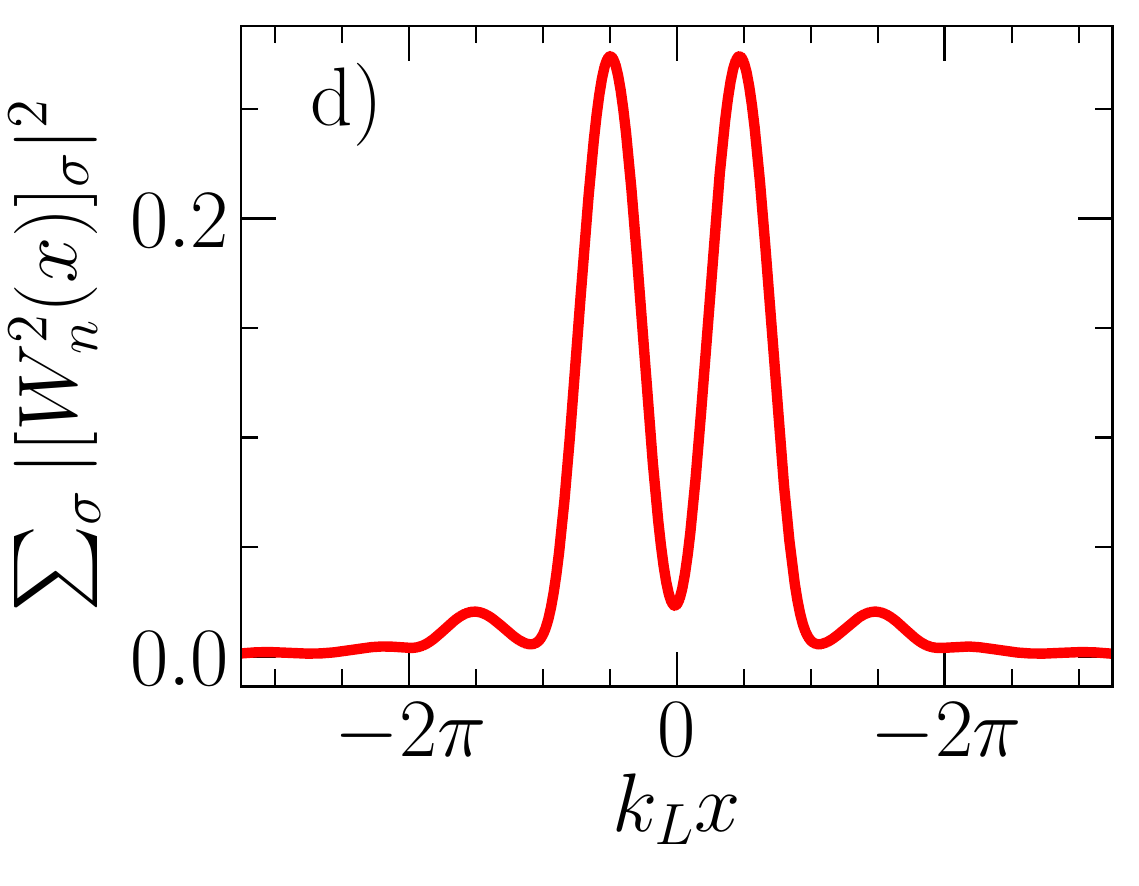}
    \caption{Sums of amplitudes moduli squared of all components $e, g_1, g_2, g_3$ of Wannier functions. Panels a), c) are for lowest band and b), d) first excited band and for $\Omega_i = (5000, 2000, 500)E_R$; a), b) $a=0.435$, c), d) $a=1.3$. Grey, dotted lines depict the same Wannier functions but in an adjacent lattice site. Orthogonality of heavily overlapping Wannier functions is possible due to the interplay of phases of individual bare atomic components  components $g_1,g_2,g_3$.
    }
    \label{fig:wanniery}
\end{figure}

{In the light of} highly non-standard relations between $J_1,J_2,J_3$ hopping amplitudes [Section~\ref{subsec:bands:tightBinding}]  it is instructive to inspect spatial profiles of  Wannier functions  of the tripod system. Again assume $\vb{\Omega}=(5000E_R,2000E_R,500E_R)$ as an example. Figure~\ref{fig:wanniery} shows the total density $||W_n^\alpha||^2$ for the Wannier functions. Panels a) and c) show the Wannier function for the lowest dark state band for $a=0.435$ and $a=1.3$ respectively. The notable feature is a non-vanishing overlap of densities of neighbouring Wannier functions
(they remain of course orthogonal to each other). For a large $a$ additional modulation shows indicating poorer confinement which corresponds well with large values of long range hopping $J_{i>3}$. 
For the first excited band panels Fig.~\ref{fig:wanniery}b) and d) show again $a=0.435$ and $a=1.3$ cases. The second band shows Wannier functions that are bimodal and have a total width $\sim \lambda_L$. Despite that the $\lambda_L/2$-displaced Wannier functions are mutually orthogonal. This is possible only because the Wannier functions can alter decomposition into separate  $g_1,g_2,g_3$ in a position dependent way. When calculating the inner product of $W_n^\alpha$ and $W_{n+1}^\alpha$ the result is zero only after the summation over $\sigma\in {g_1,g_2,g_3}$. This is not possible in a scalar Wannier function for a standard optical potential.

 Standard integrals describing two-particle interaction in Hubbard-type models are often of the type:
\begin{equation}
    U=\int {\bar{W}}_n(x) {\bar W}_m(x')V(x-x') W_o(x') W_p(x) \textrm{d}x\textrm{d}x' 
\end{equation}
For the first excited band the Wannier functions $W_n$ and $W_{n+1}$ significantly overlap allowing the above integral to yield large value of $U$ even for non-onsite processes. This potentially makes construction of discrete models with long-range interaction much easier without need to use, e.g. dipolar interactions.
This correlates well with with non-standard relations between $J_1^\alpha,J_2^\alpha,J_3^\alpha$ which allow also for long range hopping.

\section{Summary and outlooks}
\label{sec:summary}
In this work a four-level system in tripod configuration hosting two dark state is presented. The position-dependence of the dark state is set by position-dependence of Rabi frequencies $\Omega_i(x)$'s. This constraint creates periodic gauge-field like potentials that give rise to a band structure with well separated bands. In contrast to the conceptually similar case of three-level system in $\Lambda$ configuration, the lifetime of the gas populating the bands can be substantially increased. The controlling parameter is $a$ -- the phase difference between two lasers implementing the $\Omega_1$ and $\Omega_2$ couplings.
The band structure is characterized by a highly-nonstandard relations between hopping amplitudes to nearest-, next-nearest-, next-next-nearest-neighbor lattice sites, allowing for efficient long range hopping. This is reflected in non-standard shape of Wannier functions for the appropriate band that are supported at two neighboring unit cells.

When this work was close to completion we became aware of a recent preprint of E. Gvozdiovas, P. Rackauskas and G. Juzeliunas 
\cite{Gvozdiovas21} that also considered the similar tripod configuration for optical lattice creation.  We believe that
our choice of parameters minimizes the population of bright states (which makes such a lattice more stable than for the choice in~\cite{Gvozdiovas21}).

\section{Acknowledgements}

We acknowledge support by National Science Centre grant 2019/35/B/ST2/00838 (P.K. and M.Ł.) and
2017/25/Z/ST2/03029 (J.Z.). 

\appendix

\section{A different gauge choice}

It is natural to ask how other choices of the dark state basis compare to the one discussed above. 

A general different possible basis choice for the dark state subspace is given by a position-dependent two dimensional unitary transformation 
\begin{equation}
(\ket{D_1(x)},\ket{D_2(x)})\to U_2(x)(\ket{D_1(x)},\ket{D_2(x)}),
\end{equation}
by angle $\alpha(x)$. 
Such a transformation preserves the overall Hamiltonian form with $A(x)$ transformed as: 
\begin{equation} \label{eq:A_gauge}
    A \rightarrow U(x) A U^\dagger(x) - i\hbar \pdv{U(x)}{x} U^\dagger(x).
\end{equation}
where $U(x)$ is a four-dimensional extension of $U_2(x)$ including the transformation of the bright states (by phase factors).
To preserve $A_{ii}=0,i<3$ we can use $U_2(x)=\exp[i \sigma_y \alpha(x) ]$. Extra freedom to pick $\alpha(x)$ does not unfortunately allow for nullification of non-diagonal terms of $A(x)$ [see Appendix \ref{app:alpha}].

\begin{figure}[!t]
    \centering
    \includegraphics[width=0.9\linewidth]{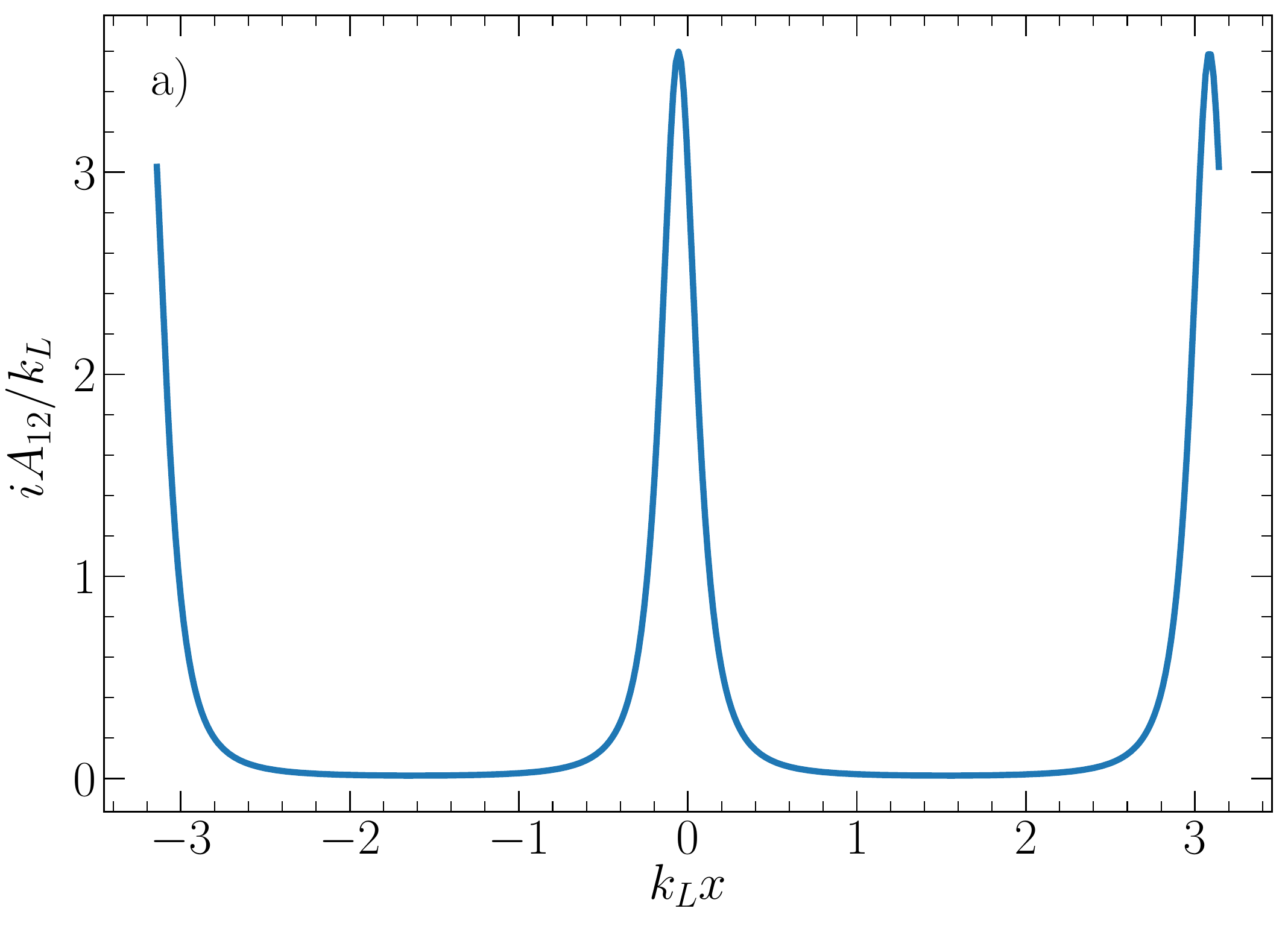}
    \includegraphics[width=0.9\linewidth]{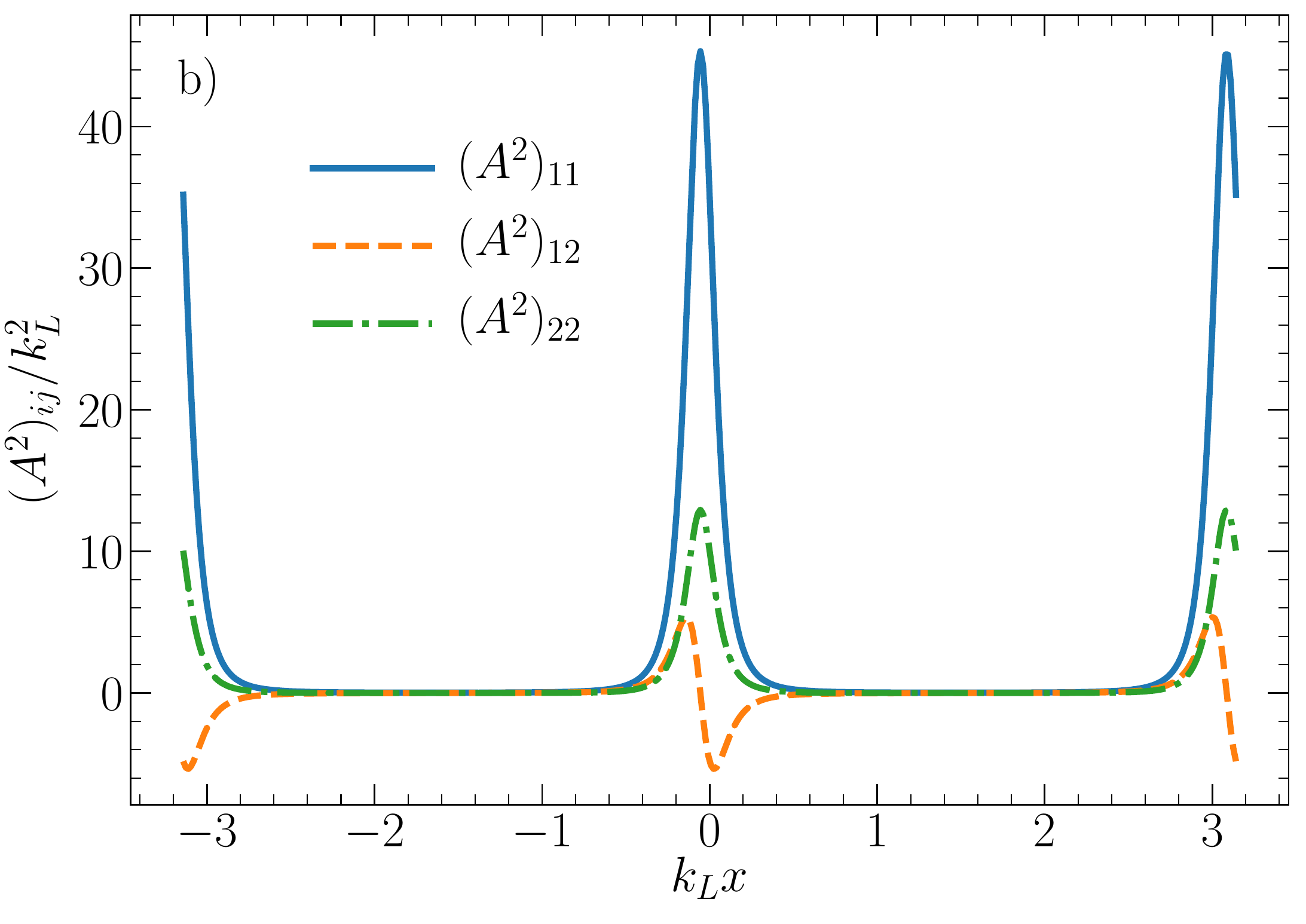}
    \includegraphics[width=0.9\linewidth]{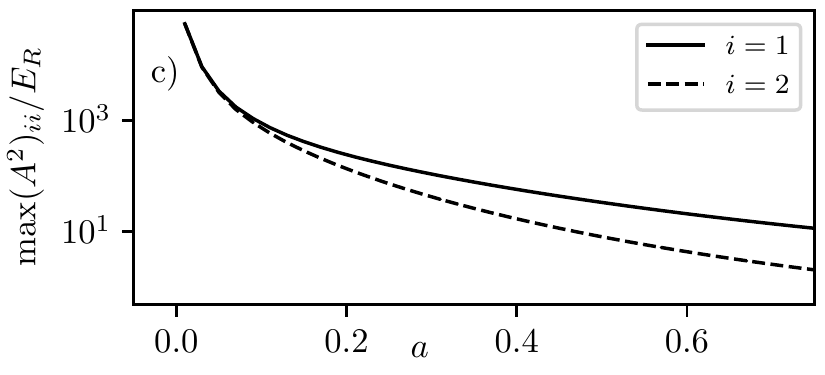}
\caption{The spatial dependence of elements of matrices $A(x)$ and $A^2(x)$, determined assuming problematic dark state definition in Eqs.~\eqref{eq:badD1} and \eqref{eq:badD2} for $\Omega_1 : \Omega_2 : \Omega_3 = 50:20:5$, $a=0.435$. Panel a) shows $A_{12}(x)$, the only non-zero element of $A$ as given by~\eqref{eq:eqAbad}. Panel b) shows elements of $A^2(x)$:  $(A^2)_{11}$,  $(A^2)_{22}$ and  $(A^2)_{12}$. Panel c) shows dependence of peak height in $(A^2)_{11}$,  $(A^2)_{22}$ on $a$ and divergent behaviour at $a\to 0$.
}
    \label{fig:A22}
\end{figure}

Features such as the height and location of $(A^2)_{11},(A^2)_{22}$ peaks and spatial dependence of terms $(A^2)_{12},(A^2)_{22}$ are strongly gauge-dependent. To illustrate this we consider an alternative gauge choice. Instead of using dark states in Eq.~\eqref{eq:bestD1} and~\eqref{eq:bestD2} let us use Eq.~\eqref{eq:badD1} and ~\eqref{eq:badD2} as a basis for determination of potential $A(x)$. Figure~\ref{fig:A22} shows the gauge potentials in that case. Due to analytical simplicity of Eq.~\eqref{eq:badD1} and ~\eqref{eq:badD2}, one can work out the formulas for the $A(x)$. They are:

\begin{align}
    A_{11}(x)&=A_{22}(x)=0 \nonumber \\
    A_{12}(x) &= -i k_L \frac{\Omega_1 \Omega_2 \Omega_3 \sin a}{\Xi_1^2(x) \Xi_2(x)}, \nonumber\\
    (A^2)_{11}(x) &= k_L^2 \qty(\frac{\Omega_1 \Omega_2 \sin a}{\Xi_1^2(x)})^2, \nonumber\\
    (A^2)_{12}(x) &= -k_L^2 \frac{\Omega_1 \Omega_2 \Omega_3 \chi(x) \sin a}{\Xi_1^2(x)\Xi_2^3(x)}, \nonumber\\
    (A^2)_{22}(x) &= k_L^2 \qty[ \qty(\frac{\Omega_1 \Omega_2 \Omega_3 \sin a}{\Xi_1^2(x)\Xi_2(x)})^2 + \qty(\frac{\Omega_3 \chi(x)}{\Xi_1(x)\Xi_2^2(x)})^2 ], \label{eq:eqAbad}
\end{align}
where
\begin{align}
    2\Xi_1^2(x) &=\left(\Omega_1^2+\Omega_2^2\right)-C \cos(2k_Lx+b)\nonumber \\
    \Xi_2^2(x) &= \Omega_3^2 + \Xi_1^2(x), \nonumber\\
    2\chi(x) &= C\sin(2k_L x + b),\nonumber\\
    C &= \sqrt{\Omega_1^4+\Omega_2^4+2\Omega_1^2\Omega_2^2\cos 2a},\nonumber\\
    \sin b &= \frac{\Omega_2^2\sin 2a}{C}.\label{eq:eqXi}
\end{align}

In this case the gauge potentials differ qualitatively from the ones in Fig.~\ref{fig:A}. First the potential maxima of $(A^2)_{11}$ and $(A^2)_{22}$ coincide and $(A^2)_{22}$ is nearly zero far from potential peaks.  In the gauge choice defined by Eq.~\eqref{eq:bestD1} and~\eqref{eq:bestD2} the potential $(A^{2})_{11}$ features a series of narrow peaks that are located between the peaks of $(A^{2})_{22}$, shifted by $\lambda_L/4$. 

Another distinct feature of the choice in Eq.~\eqref{eq:badD1} and ~\eqref{eq:badD2} is divergence of height of peaks of $(A^2)_{11}$ as $a\to 0$  [see Fig.~\ref{fig:A22}c)]. This is in stark contrast to the case of Fig.~\ref{fig:A} where $(A^2)_{11}\to 0$. This straightforwardly follows from Eq.~\eqref{eq:badD1}. This is because for $x=0$ and $x=-a$ the $|D_1(x)\rangle$ is respectively $|g_2\rangle$ and $|g_1\rangle$. This implies rapid variation of $|D_1(x)\rangle$ and in turn divergent $A^2_{11}(x)$ -- see Eq.~\eqref{eq:proofA2}.

\section{Failure to zero \texorpdfstring{$A_{12}$}{Lg} by the gauge transformation}
\label{app:alpha}

Using the gauge freedom, one could hope to vanish all elements of $A_{KL}$ in the dark subspace. Then, the dark-subspace Hamiltonian would simplify to
\begin{equation} \label{eq:ham_simple}
    H = \frac{P^2}{2m} + \frac{(A^2)_{2\times2}}{2m}.
\end{equation}
We consider a system, where $\tilde{\Delta} = 0$ and $\Omega_i$ are real giving $A_{11}=A_{22} = 0$. We choose an arbitrary, position dependent basis $\vb{d}_1(x)$, $\vb{d}_2(x)$. The convenient choice is
\begin{align} \label{eq:d}
    \vb{d}_1(x) &= \vu{\Omega} \cp \frac{\vu{\Omega}'}{\norm{\vu{\Omega}'}}, \\
    \vb{d}_2(x) &= -\frac{\vu{\Omega}'}{\norm{\vu{\Omega}'}}, \notag
\end{align}
where $\vu{\Omega} = \vb{\Omega} / \norm{\vb{\Omega}}$. The equation $A_{12} = 0$ implies 
\begin{align} \label{eq:A12_cond}
    \vb{D}_1' \vdot \vb{\Omega} &= 0, \\
    \vb{D}_2' \vdot \vb{\Omega} &= 0. \notag
\end{align}
We see, that there is still residual gauge freedom -- we can rotate the basis by any position-independent rotation matrix. This freedom is well captured when we reformulate our problem.  We express $\vb{D}_i$ as a position-dependent rotation of $\vb{d}_i$:
\begin{equation} \label{eq:D_d_rot}
    \begin{pmatrix}
        \vb{D}_1(x) \\
        \vb{D}_2(x)
    \end{pmatrix}
    =
    \begin{pmatrix}
        \cos(\alpha(x)) & -\sin(\alpha(x)) \\
        \sin(\alpha(x)) & \cos(\alpha(x))
    \end{pmatrix}
    \begin{pmatrix}
        \vb{d}_1(x) \\
        \vb{d}_2(x)
    \end{pmatrix}.
\end{equation}
After some algebra, we can write the final equation for $\alpha(x)$:
\begin{equation} \label{eq:alpha}
    \alpha'(x) = \frac{\vu{\Omega} \vdot \qty(\vu{\Omega}' \cp \vu {\Omega}'')}{\norm{\vu{\Omega}'}^2}
    = \det\qty(\vu{\Omega}, \vu{\Omega}', \vu {\Omega}'') / {\norm{\vu{\Omega}'}^2}.
\end{equation}
The solution is unique up to a constant, which reflects the residual gauge freedom. In this gauge
\begin{equation} \label{eq:A_0}
    A = \frac{i\hbar}{\sqrt{2}}\norm{\vu{\Omega}'}
    \begin{pmatrix}
        0 & 0 & \sin \alpha & \sin \alpha  \\
        0 & 0 & -\cos \alpha & -\cos \alpha \\
        -\sin \alpha & \cos \alpha & 0 & 0 \\
        -\sin \alpha & \cos \alpha & 0 & 0 \\
    \end{pmatrix},
\end{equation}
and
\begin{equation} \label{eq:A2_0}
    A^2 = \hbar^2 \norm{\vu{\Omega}'}^2
    \begin{pmatrix}
        \sin^2 \alpha & \sin \alpha \cos \alpha & 0 & 0 \\
        \sin \alpha \cos \alpha & \cos^2 \alpha & 0 & 0 \\
        0 & 0 & 1/2 & 1/2 \\
        0 & 0 & 1/2 & 1/2
    \end{pmatrix}.
\end{equation}
There is, however, one problem -- although $\alpha'(x)$ is essentially periodic for periodic $\vb{\Omega}$, the solution $\alpha(x)$ may not be, when $\int_\text{period} \dd{x} \alpha'(x) \neq n \pi,n\in\mathbb{Z}$, . Then, for $\vb{\Omega}(x) = \vb{\Omega}(x + \lambda)$, equation~\eqref{eq:D_d_rot} implies that $\vb{D}_1(x)$ and $\vb{D}_2(x)$ are aperiodic functions of $x$. This  aperiodicity translates back to remaining terms of $A(x)$ via~\eqref{eq:A_0} which makes it impossible to directly apply Bloch theory.


\begin{thebibliography}{30}%
\makeatletter
\providecommand \@ifxundefined [1]{%
 \@ifx{#1\undefined}
}%
\providecommand \@ifnum [1]{%
 \ifnum #1\expandafter \@firstoftwo
 \else \expandafter \@secondoftwo
 \fi
}%
\providecommand \@ifx [1]{%
 \ifx #1\expandafter \@firstoftwo
 \else \expandafter \@secondoftwo
 \fi
}%
\providecommand \natexlab [1]{#1}%
\providecommand \enquote  [1]{``#1''}%
\providecommand \bibnamefont  [1]{#1}%
\providecommand \bibfnamefont [1]{#1}%
\providecommand \citenamefont [1]{#1}%
\providecommand \href@noop [0]{\@secondoftwo}%
\providecommand \href [0]{\begingroup \@sanitize@url \@href}%
\providecommand \@href[1]{\@@startlink{#1}\@@href}%
\providecommand \@@href[1]{\endgroup#1\@@endlink}%
\providecommand \@sanitize@url [0]{\catcode `\\12\catcode `\$12\catcode
  `\&12\catcode `\#12\catcode `\^12\catcode `\_12\catcode `\%12\relax}%
\providecommand \@@startlink[1]{}%
\providecommand \@@endlink[0]{}%
\providecommand \url  [0]{\begingroup\@sanitize@url \@url }%
\providecommand \@url [1]{\endgroup\@href {#1}{\urlprefix }}%
\providecommand \urlprefix  [0]{URL }%
\providecommand \Eprint [0]{\href }%
\providecommand \doibase [0]{http://dx.doi.org/}%
\providecommand \selectlanguage [0]{\@gobble}%
\providecommand \bibinfo  [0]{\@secondoftwo}%
\providecommand \bibfield  [0]{\@secondoftwo}%
\providecommand \translation [1]{[#1]}%
\providecommand \BibitemOpen [0]{}%
\providecommand \bibitemStop [0]{}%
\providecommand \bibitemNoStop [0]{.\EOS\space}%
\providecommand \EOS [0]{\spacefactor3000\relax}%
\providecommand \BibitemShut  [1]{\csname bibitem#1\endcsname}%
\let\auto@bib@innerbib\@empty
\bibitem [{\citenamefont {Jaksch}\ \emph {et~al.}(1998)\citenamefont {Jaksch},
  \citenamefont {Bruder}, \citenamefont {Cirac}, \citenamefont {Gardiner},\
  and\ \citenamefont {Zoller}}]{Jaksch1998}%
  \BibitemOpen
  \bibfield  {author} {\bibinfo {author} {\bibfnamefont {D.}~\bibnamefont
  {Jaksch}}, \bibinfo {author} {\bibfnamefont {C.}~\bibnamefont {Bruder}},
  \bibinfo {author} {\bibfnamefont {J.~I.}\ \bibnamefont {Cirac}}, \bibinfo
  {author} {\bibfnamefont {C.~W.}\ \bibnamefont {Gardiner}}, \ and\ \bibinfo
  {author} {\bibfnamefont {P.}~\bibnamefont {Zoller}},\ }\href@noop {}
  {\bibfield  {journal} {\bibinfo  {journal} {Physical Review Letters}\
  }\textbf {\bibinfo {volume} {81}},\ \bibinfo {pages} {3108} (\bibinfo {year}
  {1998})}\BibitemShut {NoStop}%
\bibitem [{\citenamefont {Greiner}\ \emph {et~al.}(2002)\citenamefont
  {Greiner}, \citenamefont {Mandel}, \citenamefont {Esslinger}, \citenamefont
  {H{\"a}nsch},\ and\ \citenamefont {Bloch}}]{Greiner2002}%
  \BibitemOpen
  \bibfield  {author} {\bibinfo {author} {\bibfnamefont {M.}~\bibnamefont
  {Greiner}}, \bibinfo {author} {\bibfnamefont {O.}~\bibnamefont {Mandel}},
  \bibinfo {author} {\bibfnamefont {T.}~\bibnamefont {Esslinger}}, \bibinfo
  {author} {\bibfnamefont {T.~W.}\ \bibnamefont {H{\"a}nsch}}, \ and\ \bibinfo
  {author} {\bibfnamefont {I.}~\bibnamefont {Bloch}},\ }\href@noop {}
  {\bibfield  {journal} {\bibinfo  {journal} {nature}\ }\textbf {\bibinfo
  {volume} {415}},\ \bibinfo {pages} {39} (\bibinfo {year} {2002})}\BibitemShut
  {NoStop}%
\bibitem [{\citenamefont {Lewenstein}\ \emph {et~al.}(2012)\citenamefont
  {Lewenstein}, \citenamefont {Sanpera},\ and\ \citenamefont
  {Ahufinger}}]{Lewenstein2012}%
  \BibitemOpen
  \bibfield  {author} {\bibinfo {author} {\bibfnamefont {M.}~\bibnamefont
  {Lewenstein}}, \bibinfo {author} {\bibfnamefont {A.}~\bibnamefont {Sanpera}},
  \ and\ \bibinfo {author} {\bibfnamefont {V.}~\bibnamefont {Ahufinger}},\
  }\href@noop {} {\emph {\bibinfo {title} {Ultracold Atoms in Optical Lattices:
  Simulating quantum many-body systems}}}\ (\bibinfo  {publisher} {Oxford
  University Press},\ \bibinfo {year} {2012})\BibitemShut {NoStop}%
\bibitem [{\citenamefont {Dutta}\ \emph {et~al.}(2015)\citenamefont {Dutta},
  \citenamefont {Gajda}, \citenamefont {Hauke}, \citenamefont {Lewenstein},
  \citenamefont {Luehmann}, \citenamefont {Malomed}, \citenamefont
  {Sowi\'{n}ski},\ and\ \citenamefont {Zakrzewski}}]{Dutta15}%
  \BibitemOpen
  \bibfield  {author} {\bibinfo {author} {\bibfnamefont {O.}~\bibnamefont
  {Dutta}}, \bibinfo {author} {\bibfnamefont {M.}~\bibnamefont {Gajda}},
  \bibinfo {author} {\bibfnamefont {P.}~\bibnamefont {Hauke}}, \bibinfo
  {author} {\bibfnamefont {M.}~\bibnamefont {Lewenstein}}, \bibinfo {author}
  {\bibfnamefont {D.-S.}\ \bibnamefont {Luehmann}}, \bibinfo {author}
  {\bibfnamefont {B.~A.}\ \bibnamefont {Malomed}}, \bibinfo {author}
  {\bibfnamefont {T.}~\bibnamefont {Sowi\'{n}ski}}, \ and\ \bibinfo {author}
  {\bibfnamefont {J.}~\bibnamefont {Zakrzewski}},\ }\href
  {http://stacks.iop.org/0034-4885/78/i=6/a=066001} {\bibfield  {journal}
  {\bibinfo  {journal} {Rep. Prog. Phys.}\ }\textbf {\bibinfo {volume} {78}},\
  \bibinfo {pages} {066001} (\bibinfo {year} {2015})}\BibitemShut {NoStop}%
\bibitem [{\citenamefont {Cooper}\ \emph {et~al.}(2019)\citenamefont {Cooper},
  \citenamefont {Dalibard},\ and\ \citenamefont {Spielman}}]{Spielman2019}%
  \BibitemOpen
  \bibfield  {author} {\bibinfo {author} {\bibfnamefont {N.}~\bibnamefont
  {Cooper}}, \bibinfo {author} {\bibfnamefont {J.}~\bibnamefont {Dalibard}}, \
  and\ \bibinfo {author} {\bibfnamefont {I.}~\bibnamefont {Spielman}},\
  }\href@noop {} {\bibfield  {journal} {\bibinfo  {journal} {Reviews of modern
  physics}\ }\textbf {\bibinfo {volume} {91}},\ \bibinfo {pages} {015005}
  (\bibinfo {year} {2019})}\BibitemShut {NoStop}%
\bibitem [{\citenamefont {Boada}\ \emph {et~al.}(2012)\citenamefont {Boada},
  \citenamefont {Celi}, \citenamefont {Latorre},\ and\ \citenamefont
  {Lewenstein}}]{Boada12}%
  \BibitemOpen
  \bibfield  {author} {\bibinfo {author} {\bibfnamefont {O.}~\bibnamefont
  {Boada}}, \bibinfo {author} {\bibfnamefont {A.}~\bibnamefont {Celi}},
  \bibinfo {author} {\bibfnamefont {J.~I.}\ \bibnamefont {Latorre}}, \ and\
  \bibinfo {author} {\bibfnamefont {M.}~\bibnamefont {Lewenstein}},\ }\href
  {\doibase 10.1103/PhysRevLett.108.133001} {\bibfield  {journal} {\bibinfo
  {journal} {Phys. Rev. Lett.}\ }\textbf {\bibinfo {volume} {108}},\ \bibinfo
  {pages} {133001} (\bibinfo {year} {2012})}\BibitemShut {NoStop}%
\bibitem [{\citenamefont {Celi}\ \emph {et~al.}(2014)\citenamefont {Celi},
  \citenamefont {Massignan}, \citenamefont {Ruseckas}, \citenamefont {Goldman},
  \citenamefont {Spielman}, \citenamefont {Juzeli\ifmmode~\bar{u}\else
  \={u}\fi{}nas},\ and\ \citenamefont {Lewenstein}}]{Celi14}%
  \BibitemOpen
  \bibfield  {author} {\bibinfo {author} {\bibfnamefont {A.}~\bibnamefont
  {Celi}}, \bibinfo {author} {\bibfnamefont {P.}~\bibnamefont {Massignan}},
  \bibinfo {author} {\bibfnamefont {J.}~\bibnamefont {Ruseckas}}, \bibinfo
  {author} {\bibfnamefont {N.}~\bibnamefont {Goldman}}, \bibinfo {author}
  {\bibfnamefont {I.~B.}\ \bibnamefont {Spielman}}, \bibinfo {author}
  {\bibfnamefont {G.}~\bibnamefont {Juzeli\ifmmode~\bar{u}\else
  \={u}\fi{}nas}}, \ and\ \bibinfo {author} {\bibfnamefont {M.}~\bibnamefont
  {Lewenstein}},\ }\href {\doibase 10.1103/PhysRevLett.112.043001} {\bibfield
  {journal} {\bibinfo  {journal} {Phys. Rev. Lett.}\ }\textbf {\bibinfo
  {volume} {112}},\ \bibinfo {pages} {043001} (\bibinfo {year}
  {2014})}\BibitemShut {NoStop}%
\bibitem [{\citenamefont {Suszalski}\ and\ \citenamefont
  {Zakrzewski}(2016)}]{Suszalski16}%
  \BibitemOpen
  \bibfield  {author} {\bibinfo {author} {\bibfnamefont {D.}~\bibnamefont
  {Suszalski}}\ and\ \bibinfo {author} {\bibfnamefont {J.}~\bibnamefont
  {Zakrzewski}},\ }\href {\doibase 10.1103/PhysRevA.94.033602} {\bibfield
  {journal} {\bibinfo  {journal} {Phys. Rev. A}\ }\textbf {\bibinfo {volume}
  {94}},\ \bibinfo {pages} {033602} (\bibinfo {year} {2016})}\BibitemShut
  {NoStop}%
\bibitem [{\citenamefont {Lohse}\ \emph {et~al.}(2018)\citenamefont {Lohse},
  \citenamefont {Schweizer}, \citenamefont {Price}, \citenamefont
  {Zilberberg},\ and\ \citenamefont {Bloch}}]{Lohse18}%
  \BibitemOpen
  \bibfield  {author} {\bibinfo {author} {\bibfnamefont {M.}~\bibnamefont
  {Lohse}}, \bibinfo {author} {\bibfnamefont {C.}~\bibnamefont {Schweizer}},
  \bibinfo {author} {\bibfnamefont {H.~M.}\ \bibnamefont {Price}}, \bibinfo
  {author} {\bibfnamefont {O.}~\bibnamefont {Zilberberg}}, \ and\ \bibinfo
  {author} {\bibfnamefont {I.}~\bibnamefont {Bloch}},\ }\href {\doibase
  10.1038/nature25000} {\bibfield  {journal} {\bibinfo  {journal} {Nature}\
  }\textbf {\bibinfo {volume} {553}},\ \bibinfo {pages} {55} (\bibinfo {year}
  {2018})}\BibitemShut {NoStop}%
\bibitem [{\citenamefont {{\L}{\k{a}}cki}\ \emph {et~al.}(2016)\citenamefont
  {{\L}{\k{a}}cki}, \citenamefont {Baranov}, \citenamefont {Pichler},\ and\
  \citenamefont {Zoller}}]{Lacki2016}%
  \BibitemOpen
  \bibfield  {author} {\bibinfo {author} {\bibfnamefont {M.}~\bibnamefont
  {{\L}{\k{a}}cki}}, \bibinfo {author} {\bibfnamefont {M.}~\bibnamefont
  {Baranov}}, \bibinfo {author} {\bibfnamefont {H.}~\bibnamefont {Pichler}}, \
  and\ \bibinfo {author} {\bibfnamefont {P.}~\bibnamefont {Zoller}},\
  }\href@noop {} {\bibfield  {journal} {\bibinfo  {journal} {Physical review
  letters}\ }\textbf {\bibinfo {volume} {117}},\ \bibinfo {pages} {233001}
  (\bibinfo {year} {2016})}\BibitemShut {NoStop}%
\bibitem [{\citenamefont {Jendrzejewski}\ \emph {et~al.}(2016)\citenamefont
  {Jendrzejewski}, \citenamefont {Eckel}, \citenamefont {Tiecke}, \citenamefont
  {Juzeli{\=u}nas}, \citenamefont {Campbell}, \citenamefont {Jiang},\ and\
  \citenamefont {Gorshkov}}]{Gorshkov2016}%
  \BibitemOpen
  \bibfield  {author} {\bibinfo {author} {\bibfnamefont {F.}~\bibnamefont
  {Jendrzejewski}}, \bibinfo {author} {\bibfnamefont {S.}~\bibnamefont
  {Eckel}}, \bibinfo {author} {\bibfnamefont {T.}~\bibnamefont {Tiecke}},
  \bibinfo {author} {\bibfnamefont {G.}~\bibnamefont {Juzeli{\=u}nas}},
  \bibinfo {author} {\bibfnamefont {G.}~\bibnamefont {Campbell}}, \bibinfo
  {author} {\bibfnamefont {L.}~\bibnamefont {Jiang}}, \ and\ \bibinfo {author}
  {\bibfnamefont {A.}~\bibnamefont {Gorshkov}},\ }\href@noop {} {\bibfield
  {journal} {\bibinfo  {journal} {Physical Review A}\ }\textbf {\bibinfo
  {volume} {94}},\ \bibinfo {pages} {063422} (\bibinfo {year}
  {2016})}\BibitemShut {NoStop}%
\bibitem [{\citenamefont {Larson}\ and\ \citenamefont
  {Martikainen}(2008)}]{Larson2008}%
  \BibitemOpen
  \bibfield  {author} {\bibinfo {author} {\bibfnamefont {J.}~\bibnamefont
  {Larson}}\ and\ \bibinfo {author} {\bibfnamefont {J.-P.}\ \bibnamefont
  {Martikainen}},\ }\href@noop {} {\bibfield  {journal} {\bibinfo  {journal}
  {Physical Review A}\ }\textbf {\bibinfo {volume} {78}},\ \bibinfo {pages}
  {063618} (\bibinfo {year} {2008})}\BibitemShut {NoStop}%
\bibitem [{\citenamefont {Larson}\ and\ \citenamefont
  {Martikainen}(2009)}]{Larson2009}%
  \BibitemOpen
  \bibfield  {author} {\bibinfo {author} {\bibfnamefont {J.}~\bibnamefont
  {Larson}}\ and\ \bibinfo {author} {\bibfnamefont {J.-P.}\ \bibnamefont
  {Martikainen}},\ }\href@noop {} {\bibfield  {journal} {\bibinfo  {journal}
  {Physical Review A}\ }\textbf {\bibinfo {volume} {80}},\ \bibinfo {pages}
  {033605} (\bibinfo {year} {2009})}\BibitemShut {NoStop}%
\bibitem [{\citenamefont {Wang}\ \emph {et~al.}(2018)\citenamefont {Wang},
  \citenamefont {Subhankar}, \citenamefont {Bienias}, \citenamefont
  {{\L}{\k{a}}cki}, \citenamefont {Tsui}, \citenamefont {Baranov},
  \citenamefont {Gorshkov}, \citenamefont {Zoller}, \citenamefont {Porto},
  \citenamefont {Rolston} \emph {et~al.}}]{Wang2018}%
  \BibitemOpen
  \bibfield  {author} {\bibinfo {author} {\bibfnamefont {Y.}~\bibnamefont
  {Wang}}, \bibinfo {author} {\bibfnamefont {S.}~\bibnamefont {Subhankar}},
  \bibinfo {author} {\bibfnamefont {P.}~\bibnamefont {Bienias}}, \bibinfo
  {author} {\bibfnamefont {M.}~\bibnamefont {{\L}{\k{a}}cki}}, \bibinfo
  {author} {\bibfnamefont {T.-C.}\ \bibnamefont {Tsui}}, \bibinfo {author}
  {\bibfnamefont {M.~A.}\ \bibnamefont {Baranov}}, \bibinfo {author}
  {\bibfnamefont {A.~V.}\ \bibnamefont {Gorshkov}}, \bibinfo {author}
  {\bibfnamefont {P.}~\bibnamefont {Zoller}}, \bibinfo {author} {\bibfnamefont
  {J.~V.}\ \bibnamefont {Porto}}, \bibinfo {author} {\bibfnamefont {S.~L.}\
  \bibnamefont {Rolston}},  \emph {et~al.},\ }\href@noop {} {\bibfield
  {journal} {\bibinfo  {journal} {Physical review letters}\ }\textbf {\bibinfo
  {volume} {120}},\ \bibinfo {pages} {083601} (\bibinfo {year}
  {2018})}\BibitemShut {NoStop}%
\bibitem [{\citenamefont {Tsui}\ \emph {et~al.}(2020)\citenamefont {Tsui},
  \citenamefont {Wang}, \citenamefont {Subhankar}, \citenamefont {Porto},\ and\
  \citenamefont {Rolston}}]{Rolston2020}%
  \BibitemOpen
  \bibfield  {author} {\bibinfo {author} {\bibfnamefont {T.-C.}\ \bibnamefont
  {Tsui}}, \bibinfo {author} {\bibfnamefont {Y.}~\bibnamefont {Wang}}, \bibinfo
  {author} {\bibfnamefont {S.}~\bibnamefont {Subhankar}}, \bibinfo {author}
  {\bibfnamefont {J.~V.}\ \bibnamefont {Porto}}, \ and\ \bibinfo {author}
  {\bibfnamefont {S.~L.}\ \bibnamefont {Rolston}},\ }\href {\doibase
  10.1103/PhysRevA.101.041603} {\bibfield  {journal} {\bibinfo  {journal}
  {Phys. Rev. A}\ }\textbf {\bibinfo {volume} {101}},\ \bibinfo {pages}
  {041603} (\bibinfo {year} {2020})}\BibitemShut {NoStop}%
\bibitem [{\citenamefont {Ruseckas}\ \emph {et~al.}(2005)\citenamefont
  {Ruseckas}, \citenamefont {Juzeli{\=u}nas}, \citenamefont {{\"O}hberg},\ and\
  \citenamefont {Fleischhauer}}]{Ruseckas2005}%
  \BibitemOpen
  \bibfield  {author} {\bibinfo {author} {\bibfnamefont {J.}~\bibnamefont
  {Ruseckas}}, \bibinfo {author} {\bibfnamefont {G.}~\bibnamefont
  {Juzeli{\=u}nas}}, \bibinfo {author} {\bibfnamefont {P.}~\bibnamefont
  {{\"O}hberg}}, \ and\ \bibinfo {author} {\bibfnamefont {M.}~\bibnamefont
  {Fleischhauer}},\ }\href@noop {} {\bibfield  {journal} {\bibinfo  {journal}
  {Physical review letters}\ }\textbf {\bibinfo {volume} {95}},\ \bibinfo
  {pages} {010404} (\bibinfo {year} {2005})}\BibitemShut {NoStop}%
\bibitem [{\citenamefont {Dalibard}\ \emph {et~al.}(2011)\citenamefont
  {Dalibard}, \citenamefont {Gerbier}, \citenamefont {Juzeli{\=u}nas},\ and\
  \citenamefont {{\"O}hberg}}]{Ohberg2011}%
  \BibitemOpen
  \bibfield  {author} {\bibinfo {author} {\bibfnamefont {J.}~\bibnamefont
  {Dalibard}}, \bibinfo {author} {\bibfnamefont {F.}~\bibnamefont {Gerbier}},
  \bibinfo {author} {\bibfnamefont {G.}~\bibnamefont {Juzeli{\=u}nas}}, \ and\
  \bibinfo {author} {\bibfnamefont {P.}~\bibnamefont {{\"O}hberg}},\
  }\href@noop {} {\bibfield  {journal} {\bibinfo  {journal} {Reviews of Modern
  Physics}\ }\textbf {\bibinfo {volume} {83}},\ \bibinfo {pages} {1523}
  (\bibinfo {year} {2011})}\BibitemShut {NoStop}%
\bibitem [{\citenamefont {Lin}\ \emph {et~al.}(2011)\citenamefont {Lin},
  \citenamefont {Jim{\'e}nez-Garc{\'i}a},\ and\ \citenamefont
  {Spielman}}]{Lin11}%
  \BibitemOpen
  \bibfield  {author} {\bibinfo {author} {\bibfnamefont {Y.-J.}\ \bibnamefont
  {Lin}}, \bibinfo {author} {\bibfnamefont {K.}~\bibnamefont
  {Jim{\'e}nez-Garc{\'i}a}}, \ and\ \bibinfo {author} {\bibfnamefont {I.~B.}\
  \bibnamefont {Spielman}},\ }\href {\doibase 10.1038/nature09887} {\bibfield
  {journal} {\bibinfo  {journal} {Nature}\ }\textbf {\bibinfo {volume} {471}},\
  \bibinfo {pages} {83} (\bibinfo {year} {2011})}\BibitemShut {NoStop}%
\bibitem [{\citenamefont {Hamner}\ \emph {et~al.}(2015)\citenamefont {Hamner},
  \citenamefont {Zhang}, \citenamefont {Khamehchi}, \citenamefont {Davis},\
  and\ \citenamefont {Engels}}]{Hamner15}%
  \BibitemOpen
  \bibfield  {author} {\bibinfo {author} {\bibfnamefont {C.}~\bibnamefont
  {Hamner}}, \bibinfo {author} {\bibfnamefont {Y.}~\bibnamefont {Zhang}},
  \bibinfo {author} {\bibfnamefont {M.~A.}\ \bibnamefont {Khamehchi}}, \bibinfo
  {author} {\bibfnamefont {M.~J.}\ \bibnamefont {Davis}}, \ and\ \bibinfo
  {author} {\bibfnamefont {P.}~\bibnamefont {Engels}},\ }\href {\doibase
  10.1103/PhysRevLett.114.070401} {\bibfield  {journal} {\bibinfo  {journal}
  {Phys. Rev. Lett.}\ }\textbf {\bibinfo {volume} {114}},\ \bibinfo {pages}
  {070401} (\bibinfo {year} {2015})}\BibitemShut {NoStop}%
\bibitem [{Note1()}]{Note1}%
  \BibitemOpen
  \bibinfo {note} {In general, the more complete treatment of losses would be
  by the Lindblad master equation approach.}\BibitemShut {Stop}%
\bibitem [{Note2()}]{Note2}%
  \BibitemOpen
  \bibinfo {note} {In this work we have used standard scipy diagonalization
  function \protect \textsc {eigs}.}\BibitemShut {Stop}%
\bibitem [{\citenamefont {Dahan}\ \emph {et~al.}(1996)\citenamefont {Dahan},
  \citenamefont {Peik}, \citenamefont {Reichel}, \citenamefont {Castin},\ and\
  \citenamefont {Salomon}}]{Dahan1996}%
  \BibitemOpen
  \bibfield  {author} {\bibinfo {author} {\bibfnamefont {M.~B.}\ \bibnamefont
  {Dahan}}, \bibinfo {author} {\bibfnamefont {E.}~\bibnamefont {Peik}},
  \bibinfo {author} {\bibfnamefont {J.}~\bibnamefont {Reichel}}, \bibinfo
  {author} {\bibfnamefont {Y.}~\bibnamefont {Castin}}, \ and\ \bibinfo {author}
  {\bibfnamefont {C.}~\bibnamefont {Salomon}},\ }\href@noop {} {\bibfield
  {journal} {\bibinfo  {journal} {Physical Review Letters}\ }\textbf {\bibinfo
  {volume} {76}},\ \bibinfo {pages} {4508} (\bibinfo {year}
  {1996})}\BibitemShut {NoStop}%
\bibitem [{\citenamefont {Kohn}(1959)}]{Kohn1959}%
  \BibitemOpen
  \bibfield  {author} {\bibinfo {author} {\bibfnamefont {W.}~\bibnamefont
  {Kohn}},\ }\href@noop {} {\bibfield  {journal} {\bibinfo  {journal} {Physical
  Review}\ }\textbf {\bibinfo {volume} {115}},\ \bibinfo {pages} {809}
  (\bibinfo {year} {1959})}\BibitemShut {NoStop}%
\bibitem [{\citenamefont {Kivelson}(1982)}]{Kivelson1982}%
  \BibitemOpen
  \bibfield  {author} {\bibinfo {author} {\bibfnamefont {S.}~\bibnamefont
  {Kivelson}},\ }\href@noop {} {\bibfield  {journal} {\bibinfo  {journal}
  {Physical Review B}\ }\textbf {\bibinfo {volume} {26}},\ \bibinfo {pages}
  {4269} (\bibinfo {year} {1982})}\BibitemShut {NoStop}%
\bibitem [{\citenamefont {Marzari}\ \emph {et~al.}(2012)\citenamefont
  {Marzari}, \citenamefont {Mostofi}, \citenamefont {Yates}, \citenamefont
  {Souza},\ and\ \citenamefont {Vanderbilt}}]{Marzari2012}%
  \BibitemOpen
  \bibfield  {author} {\bibinfo {author} {\bibfnamefont {N.}~\bibnamefont
  {Marzari}}, \bibinfo {author} {\bibfnamefont {A.~A.}\ \bibnamefont
  {Mostofi}}, \bibinfo {author} {\bibfnamefont {J.~R.}\ \bibnamefont {Yates}},
  \bibinfo {author} {\bibfnamefont {I.}~\bibnamefont {Souza}}, \ and\ \bibinfo
  {author} {\bibfnamefont {D.}~\bibnamefont {Vanderbilt}},\ }\href@noop {}
  {\bibfield  {journal} {\bibinfo  {journal} {Reviews of Modern Physics}\
  }\textbf {\bibinfo {volume} {84}},\ \bibinfo {pages} {1419} (\bibinfo {year}
  {2012})}\BibitemShut {NoStop}%
\bibitem [{\citenamefont {Mikeska}\ and\ \citenamefont
  {Kolezhuk1}(2004)}]{Mikeska}%
  \BibitemOpen
  \bibfield  {author} {\bibinfo {author} {\bibfnamefont {H.-J.}\ \bibnamefont
  {Mikeska}}\ and\ \bibinfo {author} {\bibfnamefont {A.~K.}\ \bibnamefont
  {Kolezhuk1}},\ }\href@noop {} {\bibfield  {journal} {\bibinfo  {journal}
  {Lect. Notes Phys.}\ }\textbf {\bibinfo {volume} {645}},\ \bibinfo {pages}
  {1} (\bibinfo {year} {2004})}\BibitemShut {NoStop}%
\bibitem [{\citenamefont {Diep}(2004)}]{Diep}%
  \BibitemOpen
  \bibfield  {author} {\bibinfo {author} {\bibfnamefont {H.~T.}\ \bibnamefont
  {Diep}},\ }\href@noop {} {\emph {\bibinfo {title} {Frustrated spin
  systems}}}\ (\bibinfo  {publisher} {World Scientific, Singapore},\ \bibinfo
  {year} {2004})\BibitemShut {NoStop}%
\bibitem [{\citenamefont {Eckardt}(2017)}]{Eckardt17}%
  \BibitemOpen
  \bibfield  {author} {\bibinfo {author} {\bibfnamefont {A.}~\bibnamefont
  {Eckardt}},\ }\href {\doibase 10.1103/RevModPhys.89.011004} {\bibfield
  {journal} {\bibinfo  {journal} {Rev. Mod. Phys.}\ }\textbf {\bibinfo {volume}
  {89}},\ \bibinfo {pages} {011004} (\bibinfo {year} {2017})}\BibitemShut
  {NoStop}%
\bibitem [{\citenamefont {Sacha}\ \emph {et~al.}(2012)\citenamefont {Sacha},
  \citenamefont {Targo\ifmmode~\acute{n}\else \'{n}\fi{}ska},\ and\
  \citenamefont {Zakrzewski}}]{Sacha12}%
  \BibitemOpen
  \bibfield  {author} {\bibinfo {author} {\bibfnamefont {K.}~\bibnamefont
  {Sacha}}, \bibinfo {author} {\bibfnamefont {K.}~\bibnamefont
  {Targo\ifmmode~\acute{n}\else \'{n}\fi{}ska}}, \ and\ \bibinfo {author}
  {\bibfnamefont {J.}~\bibnamefont {Zakrzewski}},\ }\href {\doibase
  10.1103/PhysRevA.85.053613} {\bibfield  {journal} {\bibinfo  {journal} {Phys.
  Rev. A}\ }\textbf {\bibinfo {volume} {85}},\ \bibinfo {pages} {053613}
  (\bibinfo {year} {2012})}\BibitemShut {NoStop}%
\bibitem [{\citenamefont {Gvozdiovas}\ \emph {et~al.}(2021)\citenamefont
  {Gvozdiovas}, \citenamefont {Rackauskas},\ and\ \citenamefont
  {Juzeliunas}}]{Gvozdiovas21}%
  \BibitemOpen
  \bibfield  {author} {\bibinfo {author} {\bibfnamefont {E.}~\bibnamefont
  {Gvozdiovas}}, \bibinfo {author} {\bibfnamefont {P.}~\bibnamefont
  {Rackauskas}}, \ and\ \bibinfo {author} {\bibfnamefont {G.}~\bibnamefont
  {Juzeliunas}},\ }\href@noop {} {\enquote {\bibinfo {title} {Optical lattice
  with spin-dependent sub-wavelength barriers},}\ } (\bibinfo {year} {2021}),\
  \Eprint {http://arxiv.org/abs/2105.15148} {arXiv:2105.15148 [quant-ph]}
  \BibitemShut {NoStop}%
\end{thebibliography}
%

\end{document}